\documentclass[aps,pra,twocolumn,10pt]{revtex4-1} 
\usepackage{amsfonts,amsmath,amssymb,graphicx}

\usepackage{color,times}

\definecolor{darkblue}{rgb}{0, 0, 0.8}
\usepackage[colorlinks=true, breaklinks=true, linkcolor=darkblue, citecolor=darkblue, urlcolor=darkblue]{hyperref}
\newcommand{\ket}[1]{|#1\rangle}
\newcommand{\bra}[1]{\langle#1|}
\newcommand{\bs}{\boldsymbol}

\begin{document}

\title{Analysis of imperfections in the coherent optical excitation \\of single atoms to Rydberg states}

\author{Sylvain de L\'es\'eleuc, Daniel Barredo, Vincent Lienhard, Antoine Browaeys, and Thierry Lahaye}
\affiliation{Laboratoire Charles Fabry, Institut d'Optique Graduate School, CNRS, \\ Universit\'e Paris-Saclay, 91127 Palaiseau Cedex, France}

\begin{abstract}
We study experimentally various physical limitations and technical imperfections that lead to damping and finite contrast of optically-driven Rabi oscillations between ground and Rydberg states of a single atom. Finite contrast is due to preparation and detection errors and we show how to model and measure them accurately. Part of these errors originates from the finite lifetime of Rydberg states and we observe its $n^3$-scaling with the principal quantum number $n$. To explain the damping of Rabi oscillations, we use simple numerical models, taking into account independently measured experimental imperfections, and show that the observed damping actually results from the accumulation of several small effects, each at the level of a few percents. We discuss prospects for improving the coherence of ground-Rydberg Rabi oscillations in view of applications in quantum simulation and quantum information processing with arrays of single Rydberg atoms.
\end{abstract}

\maketitle 

Arrays of single atoms trapped in optical tweezers and excited to Rydberg states are a promising platform for quantum simulation \cite{Browaeys2016,Labuhn2016,Bernien2017,Lienhard2017,Kim2017} and quantum information processing \cite{Saffman2016}. They combine a hyperfine qubit with demonstrated individual control and one-qubit gates with high fidelities~\cite{Wang2015,Xia2015,Wang2016}, the possibility to scale the system to large numbers of qubits~\cite{Barredo2016,Endres2016,Barredo2017} and strong interactions. Coherent ground-Rydberg Rabi oscillations have been observed in dilute gases~\cite{Deiglmayr2006,Reetz-Lamour2008}, in single atoms~\cite{Johnson2008,Zuo2009,Miroshnychenko2010,Biedermann2014} and in blockaded ensemble ``superatoms''~\cite{Dudin2012,Ebert2015,Zeiher2016}. Long coherence times of ground-Rydberg Rabi oscillations are a crucial element in the context of both quantum simulation, to accurately emulate interacting systems and study their ground-state or dynamical properties~\cite{Schauss2012,Schauss2015,Labuhn2016,Bernien2017,Lienhard2017,Guardado2017}, and quantum information processing, for the implementation of two-qubit gates. Recent experimental efforts have shown an improvement in the fidelities of two-qubit gates~\cite{Maller2015,Jau2015}, but they still remain below their theoretically predicted intrinsic fidelities~\cite{Xia2013,Petrosyan2017}, as compared to other experimental platforms such as trapped ions~\cite{Monz2011,Ballance2016} or superconducting qubits~\cite{Chow2012,Martinis2014_short,Pan2017_short}. Part of this is due to imperfections in the coherent optical excitation of single atoms to Rydberg states.

\begin{figure}[b!]
\centering
\includegraphics[width=56mm]{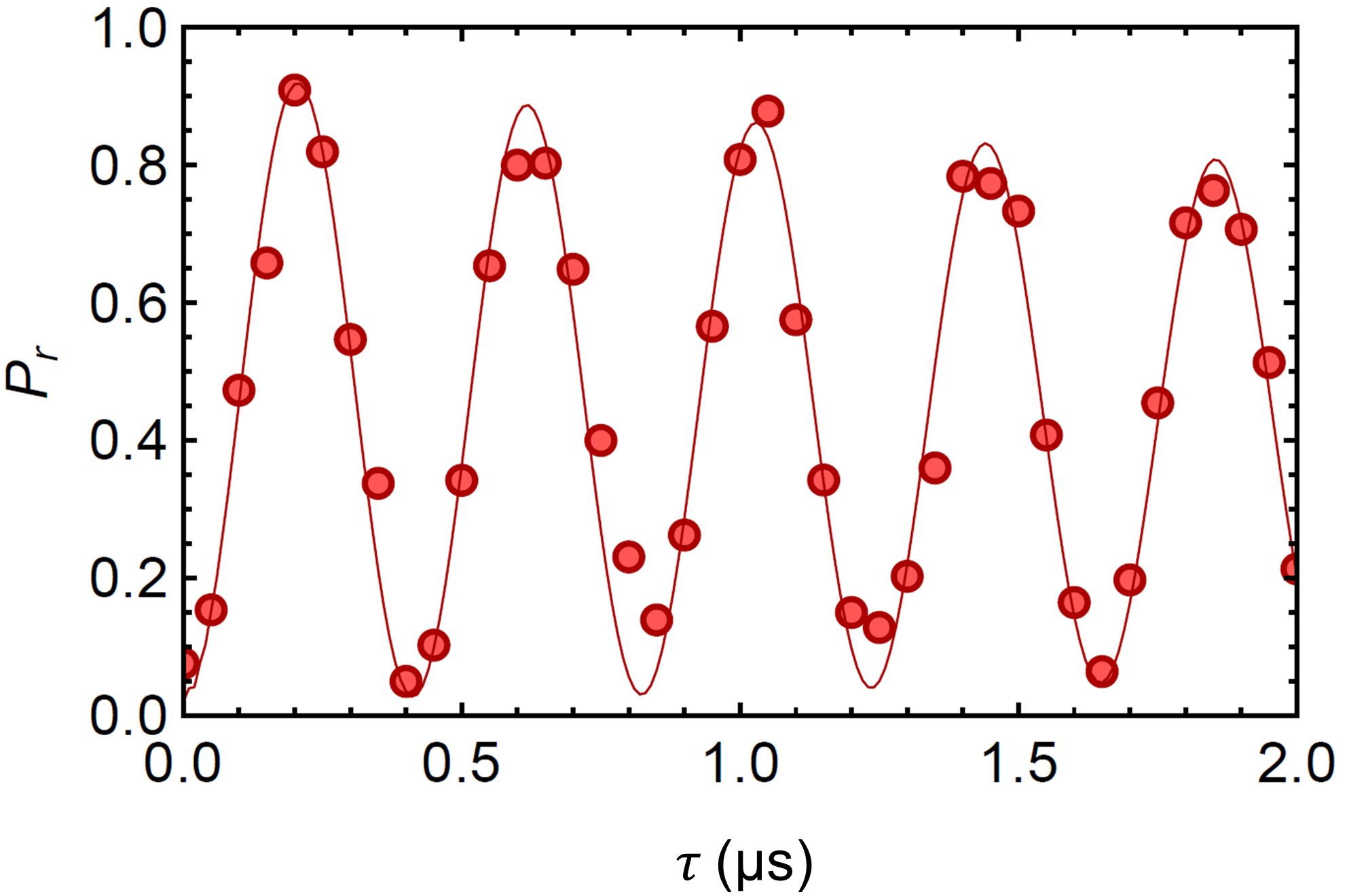}
\caption{A typical Rabi oscillation between the ground state $\ket{g}$ and the Rydberg state $\ket{r}$ (with $n=62$) when all the parameters are optimized on the experiment. The solid line is a fit by a damped sine. }
\label{fig:1}
\end{figure}

In all such experiments reported so far, one observes that the optically driven Rabi oscillations between the ground state $\ket{g}$ and the targeted Rydberg state $\ket{r}$ are damped and have a finite contrast. Figure~\ref{fig:1} gives a typical example; similar behaviors are observed in other setups~\cite{Zhang2010,Bernien2017,Zeiher2016,Biedermann2014,private}. Typical $1/e$ damping times, for a 2~MHz Rabi frequency, are about 5~$\mu{\rm s}$, much lower than the lifetime of Rydberg states, in the $\sim 200\,\mu$s range. This limits for instance the fidelity of preparation of $\ket{r}$ by a $\pi$-pulse to about 95\%. The purpose of the present study is to understand quantitatively the origins of these limitations.  

As we shall see below, they arise from the combination of several small effects due to technical imperfections in the experiment. As trying to decrease one type of imperfection may actually enhance another one, it is desirable to have a detailed understanding of all the effects at play in order to reach for the best experimental trade-off. To do so we model these imperfections as simply as possible, and compare the predictions of our models with the observed behavior, sometimes by deliberately increasing the magnitude of the deleterious effects. 

This article is organized as follows: after briefly recalling the characteristics of our setup, we review and quantify  (i) the finite efficiencies of state preparation and detection, that give a finite contrast for the Rabi flopping, but without any damping, and  (ii) the effects giving rise to damping or dephasing, among which the most significant ones are the Doppler effect, the spontaneous emission \emph{via} the intermediate state used for the two-photon excitation, and the laser phase noise. Then, combining all the effects in a global simulation, we compare with experimental results and discuss possible routes towards an improvement of the fidelities.

\section{Experimental setup}
\label{sec:exp_setup}

\begin{figure}[t!]
\centering
\includegraphics[width=80mm]{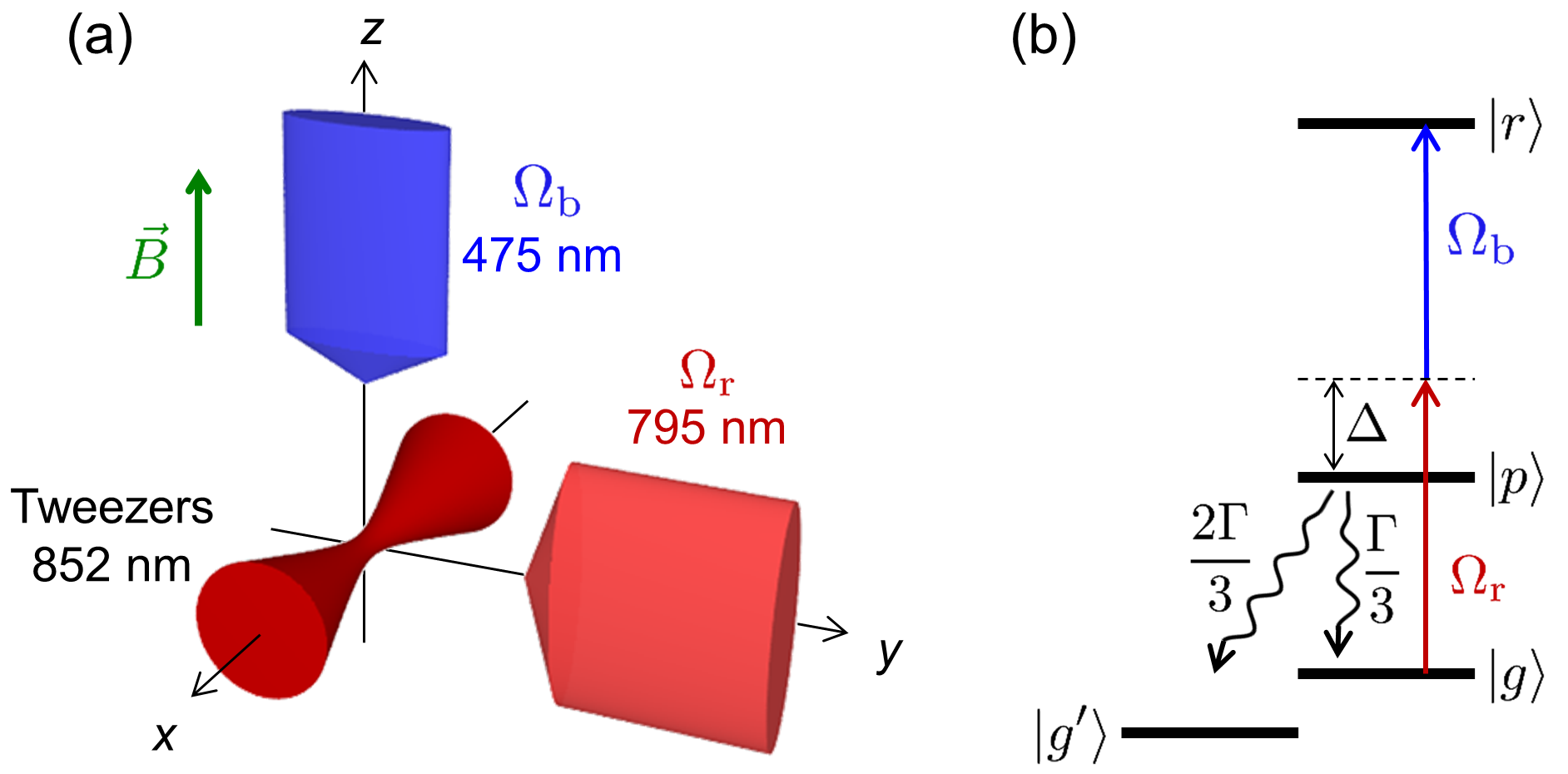}
\caption{(a): The excitation setup. The tweezers has a $1/e^2$ radius of $1.1~\mu{\rm m}$, and the blue and red beams have elliptical cross-sections, with waists $(w_x,w_y)= 24 \times 50 \;\mu{\rm m}^2$ and $(w_x,w_z)= 50\times200\;\mu{\rm m}^2$, respectively. (b): Relevant levels involved in the Rydberg excitation (see text).}
\label{fig:notations}
\end{figure}

The geometry of our experimental setup \cite{Labuhn2016} is shown in Fig.~\ref{fig:notations}(a). A Rydberg excitation sequence is as follows. We first check for the presence of a single $^{87}$Rb atom in the 1~mK-deep tweezers by shinning the molasses beams during $50$~ms and collecting the fluorescence photons on an EMCCD camera. The atom is then cooled from a temperature of $60 \, \mu$K after imaging to about $T = 30 \, \mu$K by first increasing the detuning of the molasses beam and then by adiabatic lowering of the trap depth~\cite{Tuchendler2008}. An external magnetic field along $z$ of typically 7~G is then switched on to define the quantization axis, and, after $50$~ms, we optically pump the atom into the ground state $\ket{g}=\ket{5S_{1/2},F=2,m_F=2}$. 

We then switch off the tweezers, and illuminate the atom with our Rydberg excitation lasers for a time $\tau$. As shown in Fig.~\ref{fig:notations}(b), we use a two-photon scheme (wavelengths 795~nm and 475~nm, Rabi frequencies $\Omega_{\rm r}$ and $\Omega_{\rm b}$, polarizations $\pi$ and $\sigma^+$) with a single-photon detuning $\Delta=2\pi\times 740 $~MHz from the intermediate state $\ket{p}=\ket{5P_{1/2},F'=2,m_{F'}=2}$. This results in a coherent coupling with an effective Rabi frequency $\Omega=\Omega_{\rm r}\Omega_{\rm b}/(2\Delta)$ between $\ket{g}$ and a single Rydberg Zeeman state $\ket{r}=\ket{nD_{3/2},m_J=3/2}$. With this choice of lasers polarizations, we avoid the off-resonant coupling to other Zeeman states, which would lead to dephasing. To vary $\Omega$, we tune $\Omega_{\rm r}$ (by varying the 795~nm laser power) and keep $\Omega_{\rm b}$ maximized. The latter depends on the principal quantum number and, using Autler-Townes spectroscopy~\cite{Hennrich2017}, we measured $\Omega_{\rm b}/(2\pi) = 34.8(5) \times (n^\star/60)^{-3/2}$~MHz, with the effective principal quantum number $n^\star = n - \delta_{0}$, where $\delta_0 \simeq 1.35$ is the quantum defect of $^{87}$Rb $D_{3/2}$ states.

Finally, after this excitation time $\tau$, which takes up to a few microseconds, we switch on the tweezers again. An atom in $\ket{g}$ is recaptured with high efficiency (see below), while an atom in $\ket{r}$ is repelled by the tweezers and thus lost. We then take a second fluorescence image to check for the presence of the atom. Repeating this sequence (typically 100 to 200 times) allows us to reconstruct the \emph{recapture probability}, that we denote $P_g$ as, to a first approximation, it gives the population of $\ket{g}$. The inferred Rydberg excitation probability is denoted as $P_r=1-P_g$.

\section{State preparation and detection errors}
\label{sec:deterror}

\begin{figure}[t]
\centering
\includegraphics[width=55mm]{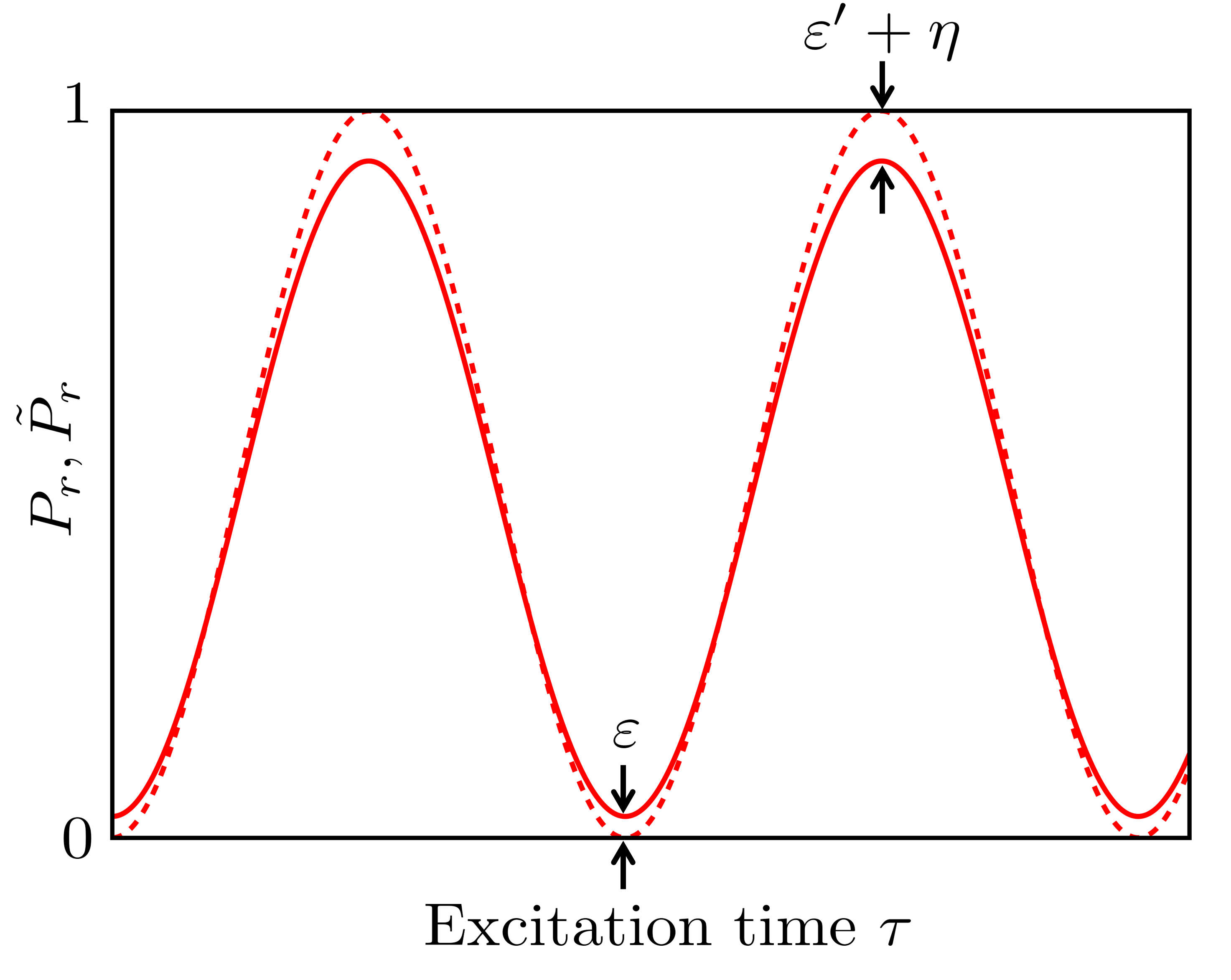}
\caption{Effect of small but finite values of $(\varepsilon,\varepsilon',\eta)$ on the measured probability $P_r$ (solid line) assuming a perfect Rabi oscillation $\tilde{P}_r$ (dashed line).}
\label{fig:eps}
\end{figure}

Even if the excitation process of an atom in $\ket{g}$ to the Rydberg state $\ket{r}$ were perfect, the measured recapture probability would not show perfectly contrasted oscillations due to state preparation and measurement errors (`SPAM' errors). We identify three different components: (a) the finite efficiency of the optical pumping leading to a preparation error with probability $\eta$, (b) the ``false positive'' errors in the Rydberg detection, as we have a probability $\varepsilon = P(r \vert g)$ to incorrectly infer that a ground-state atom was in $\ket{r}$ because it was lost, e.g. due to background-gas collisions, and (c) the ``false negative'' errors, with a probability $\varepsilon^\prime = P(g \vert r)$ to recapture a Rydberg atom which has quickly decayed back to the ground-state. 

We denote by $\tilde{P}_g$ and $\tilde{P}_r$ the actual population of the states $\ket{g}$ and $\ket{r}$ due to the evolution of the system under the excitation laser, possibly in the presence of the damping and dephasing mechanisms to be discussed in Sec.~\ref{sec:damping}. Due to the nonzero values of $(\eta, \varepsilon, \varepsilon')$, the measured probabilities of recapture $P_g$ and of loss $P_r$ are slightly altered and become:
\begin{eqnarray}
P_g&=&\eta(1-\varepsilon)+(1-\eta)(1-\varepsilon)\left[\tilde{P}_g+\varepsilon'\tilde{P}_r\right],\\
P_r&=&\eta\varepsilon+(1-\eta)\left[\varepsilon \tilde{P}_g+(1-\varepsilon'+\varepsilon\varepsilon')\tilde{P}_r\right].
\label{eq2}
\end{eqnarray}
It implies that, even if the ``real'' population $\tilde{P}_r$ undergoes a perfect Rabi oscillation $\tilde{P}_r(t)=\sin^2(\Omega \tau/2)$, the measured one $P_r(t)$ has a finite contrast. Figure \ref{fig:eps} illustrates the effect to lowest order in $(\varepsilon,\varepsilon',\eta)$. In principle, one can invert the above equations~\cite{Duan2012}, e.g. using a maximum likelihood procedure, to correct the measured populations for these errors and recover the ``real'' populations, even for many qubits~\cite{Bernien2017}. In our publications we however include these SPAM errors on the theoretically calculated populations when comparing with data~\cite{Barredo2015,Labuhn2016}. In the following, we investigate in detail the causes of those `SPAM' errors.

\paragraph{Efficiency of optical pumping.} The optical pumping into $\ket{g}$ is not entirely perfect, and we denote by $\eta$ the probability that after optical pumping, the atom is not in $\ket{g}$ but in another Zeeman or hyperfine state of $5S_{1/2}$. Measurements using microwave transitions between the two hyperfine levels of $5S_{1/2}$ allow us to estimate an upper-bound on the preparation error $\eta < 0.005$.

\paragraph{Detection errors: false positives.} The detection of the final state of the atom relies on its recapture, which ideally always occurs if the atom is in $\ket{g}$, and never occurs if it is in $\ket{r}$. However, there is a finite probability $\varepsilon$ to lose a ground-state atom during the sequence. A first source of errors are collisions of the single atom with the background gas, which gives a vacuum-limited lifetime of 20~s for an atom in the 1~mK deep tweezers without any cooling light. Integrated over the $\simeq 50$~ms required to perform the experiment (limited by eddy currents when switching off the external B-field), it amounts to an error rate of $0.3 \, \%$. During the fluorescence imaging, when the molasses beam are let on, the single atom lifetime is reduced to 8~s while the atom temperature does not increase. This could be due to the trapped atom dynamics under illumination with cooling light~\cite{Martinez-Dorantes2017b} or to residual loading from background Rb vapor. The atom can thus leave the trap during the $50$~ms fluorescence images and we estimate that this second cause of detection errors amount to $0.6 \, \%$. A third reason for losing the atom comes from its displacement, due to its finite temperature $T$, when the optical tweezers is switched off during the experimental time $\tau$, as explained in Appendix~\ref{app:epsp}. We calculate that the probability for this event remains below $1 \, \%$ for $T=30 \, \mu$K and $\tau < 6 \, \mu$s, such that by adding all three contributions we predict a false positive rate $\varepsilon < 2 \, \%$. When measuring it by repeating the experimental sequence described in Sec.~\ref{sec:exp_setup} with the traps switched off during a time $\tau \simeq 1-6 \, \mu$s but without shining the excitation beam, we indeed obtain a typical loss rate $\varepsilon \simeq 1-2$\%, in good agreement with the above estimate. 

\begin{figure}[t]
	\centering
	\includegraphics[width=80mm]{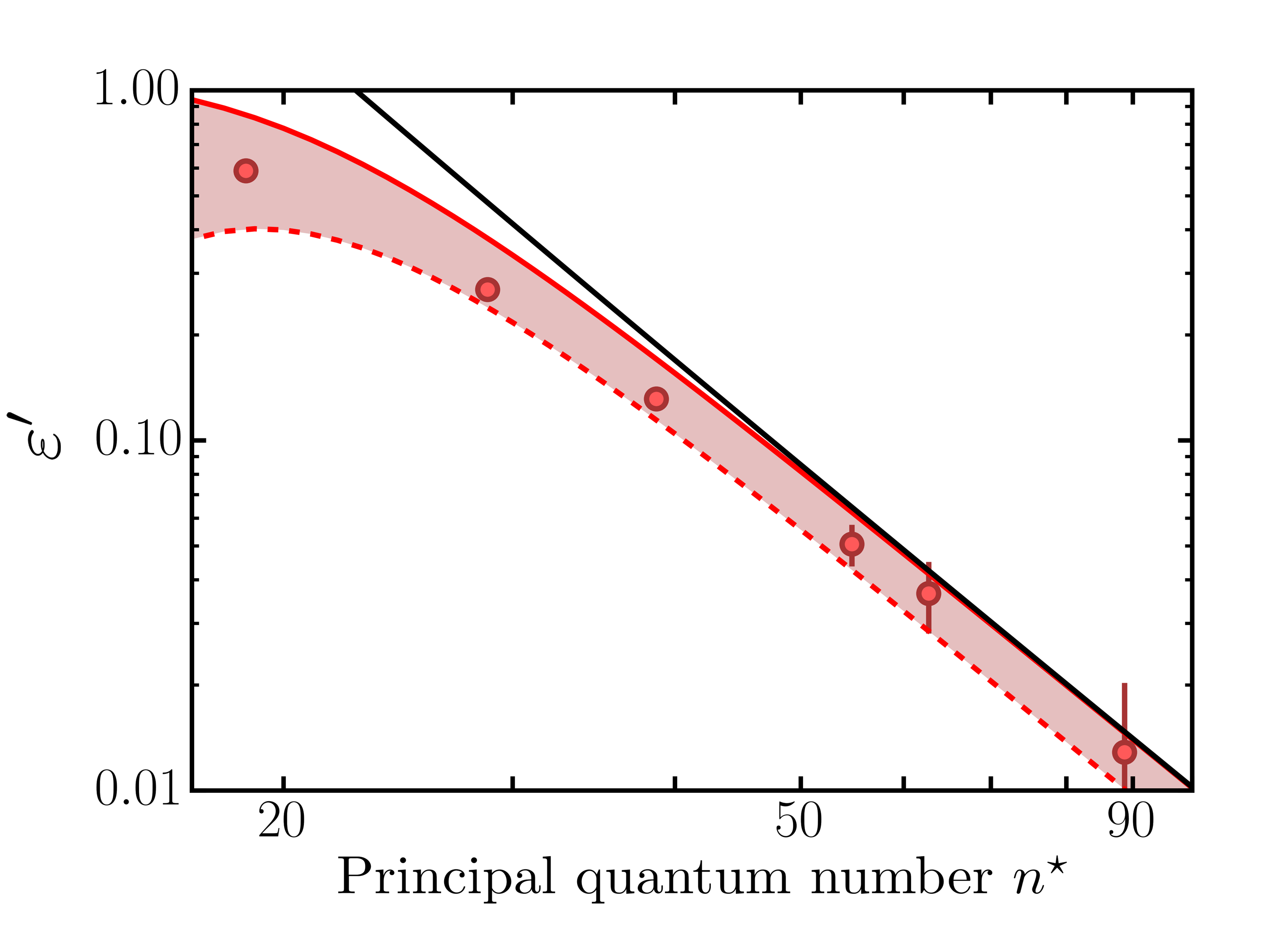}
	\caption{Detection errors $\varepsilon^\prime$ caused by the finite Rydberg state lifetime. Black solid line: $\Gamma_R t_\text{recap} $. Red solid line: Eq.~(\ref{eq:epsprime}). The red dashed line takes into account the $4 \, \mu$s push-out time to remove atoms in $\ket{g}$, see Appendix~\ref{app:epsp}. }
	\label{fig:epspr}
\end{figure}

\paragraph{Detection errors: false negatives.} Conversely, an atom that has been excited to $\ket{r}$ has a small but nonzero probability to be recaptured as it can decay back to the trapped ground state before having been expelled from the trapping region. An atom in $\ket{r}$ is repelled by the optical tweezers due to the ponderomotive force (see Appendix~\ref{app:epsp}). The effect is well captured by a measured characteristic time $t_\text{recap} = 10 \, \mu$s during which a Rydberg atom stays in the trapping region. The probability of false negative $\varepsilon^{\prime} = P(g \vert r)$ is then directly linked to the rate $\Gamma_R$ at which a Rydberg state decays to the ground-state (via low-lying excited states), which scales with the principal quantum number as $n^{-3}$. For $n > 50$, $t_\text{recap}$ is much shorter than the inverse decay rate $\Gamma_R^{-1} > 100 \, \mu\text{s}$ and we can approximate $\varepsilon^{\prime}$ by $ \Gamma_R t_\text{recap}$ (Fig.~\ref{fig:epspr}, solid black curve). For lower $n$, the approximation is not valid anymore and  $\varepsilon^{\prime}$ is given by Eq.~(\ref{eq:epsprime}) (solid red curve). We measure the detection error by (i) exciting the atom to $\ket{r}$ with probability $\tilde{P}_r$, (ii) pushing out, with unit efficiency, atoms still in $\ket{g}$~\footnote{A laser beam resonant on the ${\vert} F=2 {\rangle} {\rightarrow} {\vert} F' = 3 {\rangle}$ cycling transition at 780~nm is shone during $4 \, \mu$s (in the presence of a repumper beam). Atoms in any hyperfine and Zeeman levels of $5S_{1/2}$ are ejected with a probability better than $99.5\%$, while atoms in ${\vert} r {\rangle}$ are not affected. A similar technique was used for the detection of Rydberg states on a dark background in~\cite{Schauss2012}.}, and (iii) observing the presence or absence of the atom, which effectively measures $\varepsilon^\prime \tilde{P}_r$. Together with the measurement of $P_r$ and Eq.~(\ref{eq2}), we extract the real Rydberg fraction $\tilde{P}_r$ and the false detection error rate $\varepsilon^\prime$. The results obtained for Rydberg states ranging from $n=20$ to $90$ (disks) show the expected $n^{-3}$-scaling and are in quantitative agreement with our model. In our experiments, we use $n > 50$ Rydberg state and the false negative rate is limited to $\varepsilon^\prime < 0.05$. This error becomes more severe with tweezers schemes also trapping Rydberg states as proposed in \cite{Zhang2011}. To improve this detection method, one could consider, e.g, ionizing the Rydberg atoms by applying a strong electric field.

\section{Damping of the Rabi oscillations}
\label{sec:damping}

We now turn to effects that lead to a decreasing amplitude of the Rabi oscillation when the excitation time $\tau$ increases.

\subsection{Doppler effect}

\begin{figure}[t]
\centering
\includegraphics[width=85mm]{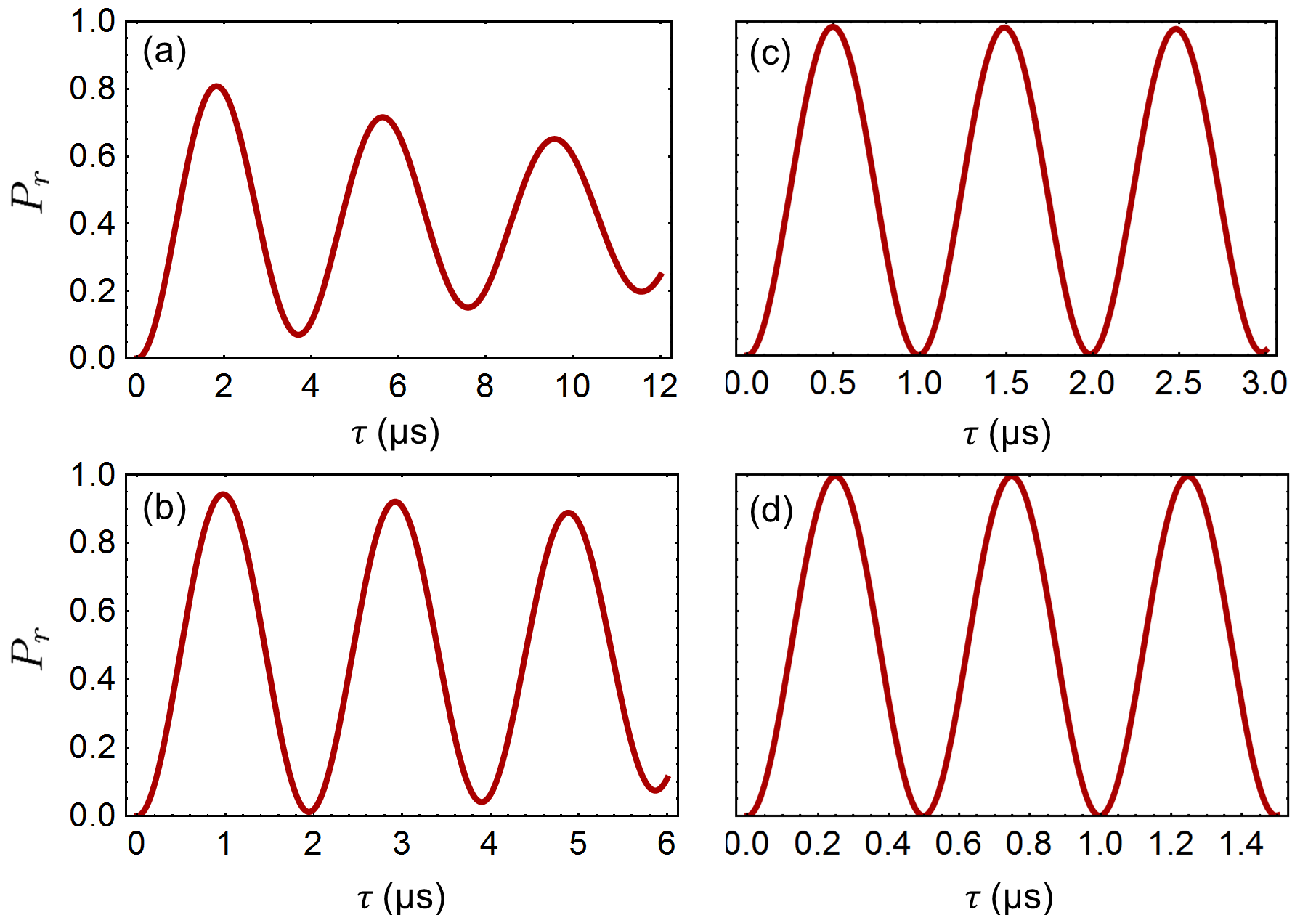}
\caption{Influence of the Doppler effect on the Rabi oscillations, for a temperature $T=30\;\mu$K and Rabi frequencies $\Omega/(2\pi)$ of 250~kHz (a), 500~kHz (b), 1~MHz (c), and 2~MHz (d).}
\label{fig:doppler}
\end{figure}

The first contribution to damping is the Doppler effect. In our setup (Fig.~\ref{fig:notations}(a)), the two excitation lasers with wavevectors ${\bs k}_{\rm r}$ and ${\bs k}_{\rm b}$ are orthogonal to each other, resulting in an effective wavevector of magnitude $k_{\rm eff}\simeq 1.5\times10^7\;{\rm m}^{-1}$. A temperature $T=30\;\mu{\rm K}$ corresponds to a one-dimensional r.m.s. velocity spread $\Delta v=\sqrt{k_{\rm B}T/m}\simeq 50\;{\rm mm/s}$. This means that for each realization of the experiment, the detuning of the excitation laser seen by the atom is a random variable with a centered Gaussian probability distribution of standard deviation $k_{\rm eff}\Delta v~\sim 2\pi\times 120$~kHz. Figure~\ref{fig:doppler} shows the calculated influence of the Doppler effect for various Rabi frequencies $\Omega/(2\pi)$. One can see that below 1~MHz, the Doppler effect is a very significant source of dephasing, while from 1~MHz up the effect is hardly noticeable. 

If we exclude more technically demanding ways to substantially decrease the Doppler effect, e.g. by using three-photon excitation~\cite{Ryabtsev2011} or reducing the temperature by Raman cooling~\cite{Kaufman2012,Thompson2013}, the above results seem to indicate that one should use high Rabi frequencies. However this is not the case, because a competing effect arises, namely spontaneous emission via the intermediate $5P_{1/2}$ state.  

\subsection{Spontaneous emission from intermediate state}
\label{ssec:spontaneous}
Although the intermediate state detuning $\Delta$ is large compared to $\Omega_{\rm r,b}$, there is still a small probability for the atom to be in the intermediate state $\ket{p}$, which has a natural linewidth $\Gamma\simeq 2\pi\times 6$~MHz. From $\ket{p}$, the atom decays back to $\ket{g}$ with a probability $1/3$, and to the other sublevels of $5S_{1/2}$ with a probability $2/3$. 

To take into account this spontaneous emission in the simplest way, we use a 4-level model for the atoms~\cite{Miroshnychenko2010}, with the states $\ket{g}$, $\ket{r}$, $\ket{p}$ and an extra state $\ket{g'}$ which accounts for all the ground-state sublevels, other than $\ket{g}$, to which the atom can decay from $\ket{p}$ (see Figure~\ref{fig:notations}b). We solve the optical Bloch equations (OBEs) for the density matrix $\rho$:
\begin{equation}
\frac{{\rm d}\rho}{{\rm d}t}=\frac{1}{i\hbar}[H,\rho]+{\cal L}[\rho]
\end{equation}
where the Hamiltonian reads, in the rotating wave approximation,
\begin{eqnarray}
H&=&\frac{\Omega_{\rm r}}{2}\left(\ket{g}\bra{p}+\ket{p}\bra{g}\right)
+\frac{\Omega_{\rm b}}{2}\left(\ket{p}\bra{r}+\ket{r}\bra{p}\right)\\
&&-\Delta \ket{p}\bra{p}-\delta\ket{r}\bra{r}.
\end{eqnarray}
Here $\delta$ is the (small) detuning from the two-photon resonance condition. The dissipator has the Lindblad form:
\begin{equation}
{\cal L}[\rho]=\sum_{i=g,g'}\frac{\Gamma_i}{2}\left(2\ket{i}\bra{p}\rho\ket{p}\bra{i}-\ket{p}\bra{p}\rho-\rho\ket{p}\bra{p}\right),
\end{equation} 
with $\Gamma_g=\Gamma/3$ and $\Gamma_{g'}=2\Gamma/3$. Here, decay of $\ket{r}$ is neglected (see Sec.~\ref{other_effects}). The recapture probability is then $1-\rho_{rr}$. 

\begin{figure}[t]
\centering
\includegraphics[width=83mm]{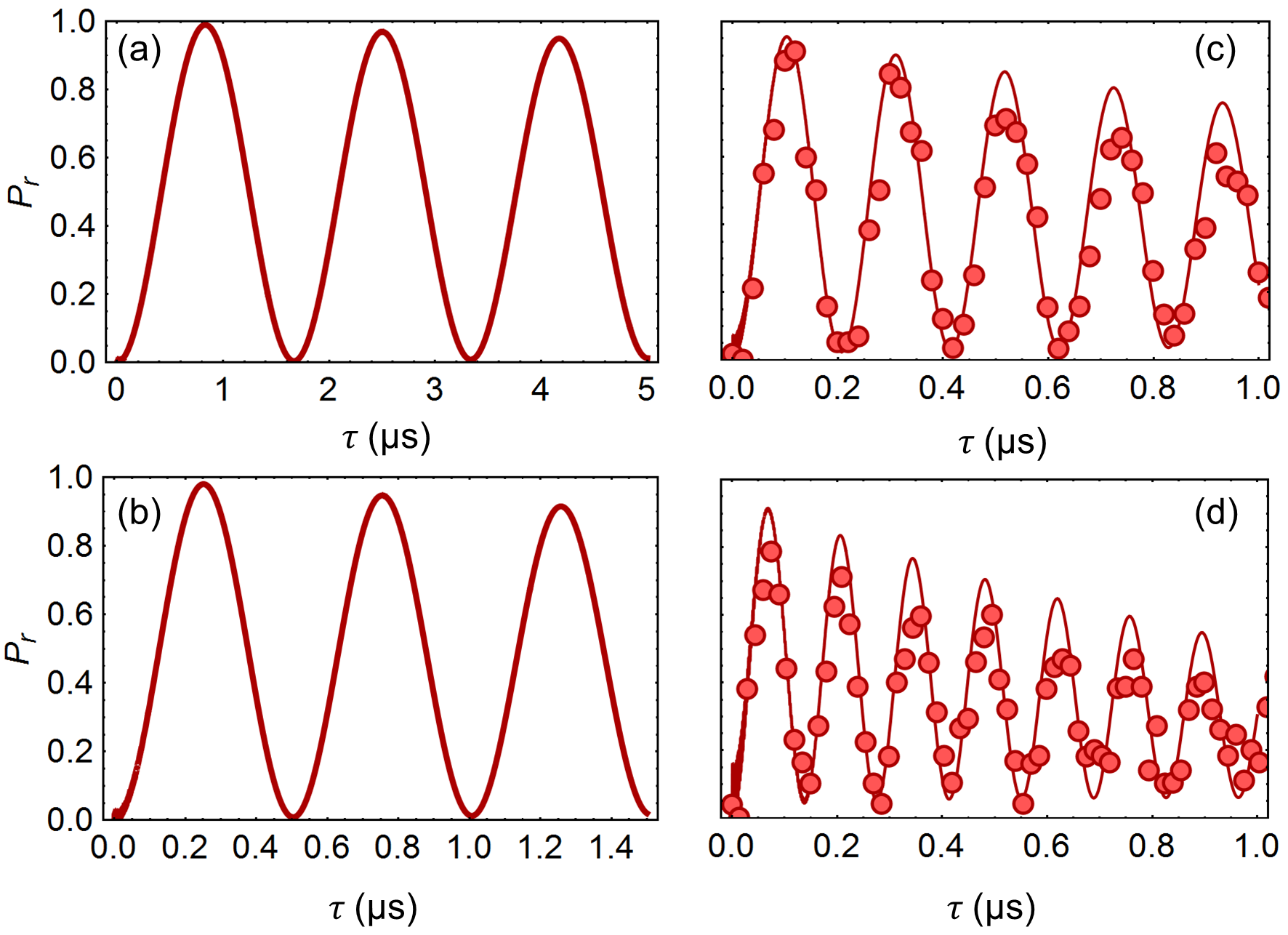}
\caption{Influence of the spontaneous emission from $\ket{p}$. (a-b): Calculated Rabi oscillation obtained by solving the OBE's for $\Delta=2\pi\times 740$~MHz, $\Omega_{\rm b}=2\pi\times 30$~MHz, and $\Omega_{\rm r}/(2\pi)= 30$~MHz (a) and 100~MHz (b). (c-d): Comparison between the simulation and experimental data (with $n = 61$), for fixed $\Omega_{\rm b}=2\pi\times 35$~MHz and $\Omega_{\rm r}= 2\pi \times 210$~MHz, but for decreasing values of the intermediate state detuning: (c), $\Delta=2\pi\times 740$~MHz, (d): $\Delta=2\pi\times 477$~MHz.}
\label{fig:spontem}
\end{figure}

Figure \ref{fig:spontem}(a,b) shows the results of such simulations for the values of $\Omega_{\rm b}$ and $\Delta$ that we generally use. The two-photon detuning $\delta$ was adjusted to compensate for the light-shifts $\left(\Omega_{\rm r}^2-\Omega_{\rm b}^2\right)/(4\Delta)$. When increasing $\Omega_{\rm r}$ we observe a stronger and stronger damping, which also shows a characteristic asymmetry: the successive maxima of $P_r$ become significantly smaller, while the minima remain quite close to zero. This is simple to understand: spontaneous emission \emph{via} $\ket{p}$ slowly optically pumps the atoms into the dark state $\ket{g'}$, and these atoms will be detected as ground-state atoms, even though, not being any more in $\ket{g}$, they have no possibility of being excited to $\ket{r}$.

In Fig.~\ref{fig:spontem}(c,d) we compare the prediction of the simulation to experimental data where we reduced $\Delta$ to enhance on purpose the effects of spontaneous emission. We get a good agreement (without adjustable parameters), giving confidence in the simple model we use. We note that the problem of spontaneous emission from an intermediate state is avoided when using a direct single-photon excitation scheme~\cite{Biedermann2014} or minimized when choosing a higher intermediate state with smaller natural linewidth (for instance, when using an ``inverted'' two photon scheme as in ~\cite{Bernien2017}, the linewidth of the intermediate $6P_{1/2}$ state is only 1.3~MHz).

\subsection{Laser phase noise}

Another important effect giving rise to damping of the oscillations is the fact that the excitation lasers have finite phase noise. We use two extended-cavity diode lasers (ECDLs), one at 795~nm, and one at 950~nm which is amplified in a tapered amplifier and frequency-doubled in a resonant cavity. Both ECDLs are locked using the Pound-Drever-Hall (PDH) technique on a high-finesse ultra-stable ULE cavity (${\cal F}\sim 20,000$ for both wavelengths). Their phases $\phi_i(t)$ ($i=795,950$) are random processes with a power spectral density  $S_{\phi_i}(f)$, where $f$ denotes the Fourier frequency. Measuring $S_{\phi_i}(f)$ directly is not an easy task, but we obtain a reasonable estimate of $S_{\phi_i}(f)$ for Fourier frequencies $f$ above acoustic frequencies, where the cavity noise is negligible, by analyzing the in-loop PDH error signal with an RF spectrum analyzer: the noise spectral density of the PDH error signal voltage allows to retrieve $S_{\phi_i}(f)$, knowing the slope of the PDH error signal at the lock point and taking into account the storage time of light in the cavity which causes a roll-off of the cavity response to frequency fluctuations (see e.g.~\cite{phdtarallo}, page~17).

\begin{figure}[t]
\centering
\includegraphics[width=80mm]{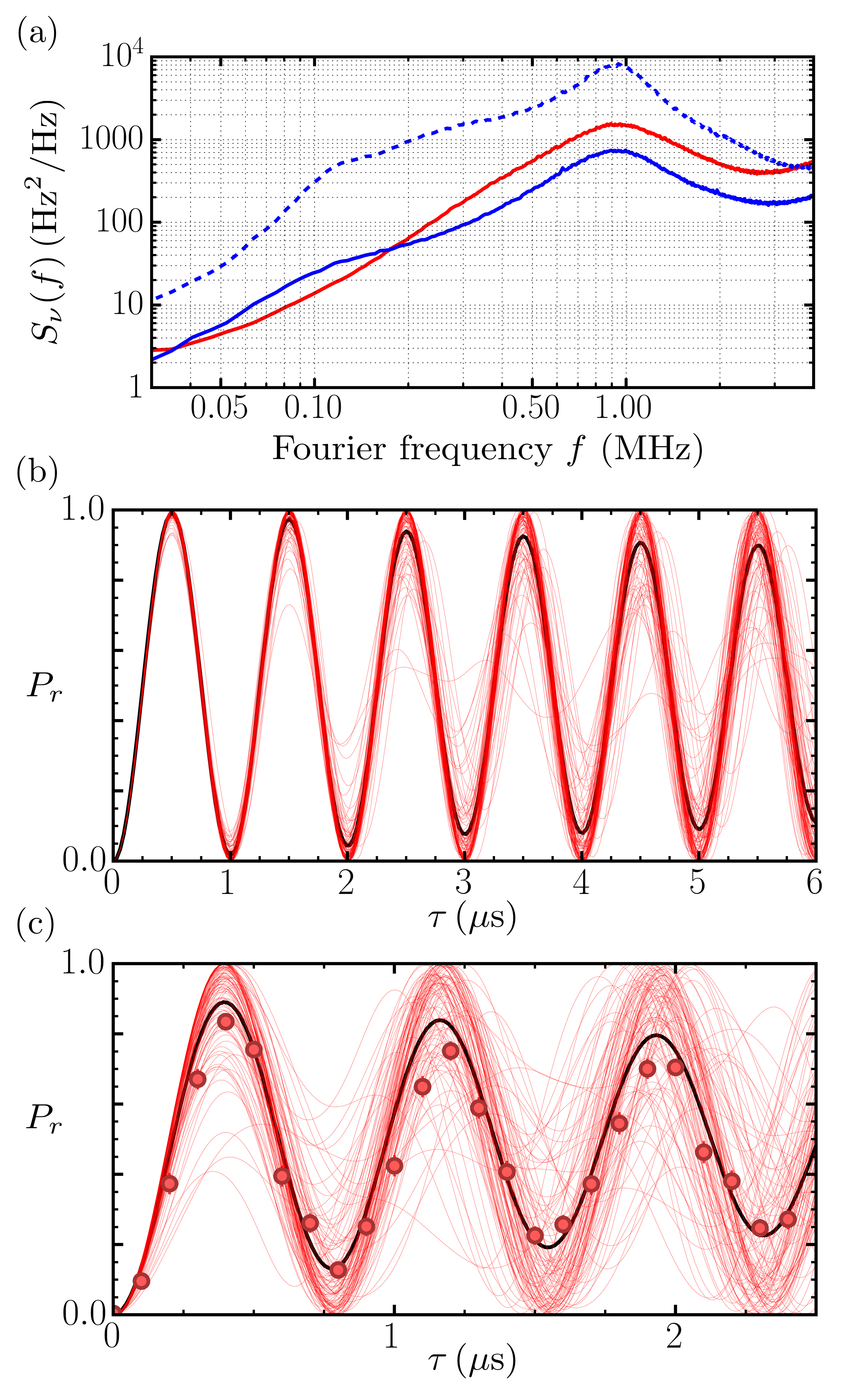}
\caption{Influence of laser phase noise. (a): Estimated spectral density $S_\nu(f)$ of the frequency noise of the 795~nm (red) and 950~nm laser (blue). Solid lines: usual noise, dashed line: enhanced noise (see text). (b): Simulated Rabi oscillation in the presence of usual phase noise. Many Rabi oscillations (thin lines), each for a given realization of the random processes $\phi_i(t)$, are averaged (thick black line). (c): With intentional extra noise added in the 950~nm laser the damping is increased, and the simulation compares well with the experimentally observed Rabi oscillation on $n = 54$ (red circles). }
\label{fig:s-nu}
\end{figure}

Figure~\ref{fig:s-nu}(a) shows the spectral density of \emph{frequency noise} $S_{\nu_i}(f)$, which is related to the phase noise spectral density by $S_{\nu_i}(f)=f^2 S_{\phi_i}(f)$~\cite{riehle2004}. The solid lines represent the typical frequency noise measured when operating the experiment. We observe a broad maximum of noise around 1~MHz, due to the limited feedback loop bandwidth, which implies that phase noise will have the highest detrimental effects for Rabi frequencies $\Omega/(2\pi)$ around this value (phase fluctuations at $2\pi f\ll \Omega$ are seen as a constant phase by the atoms and do not have any influence, while high-frequency ones are averaged out during the evolution of the atom, resulting in very little dephasing).

In order to assess quantitatively the influence of laser phase noise, rather than calculating analytically a sensitivity function~\cite{martin},  we perform a direct simulation of the dynamics of the atom in the presence of phase-fluctuating Rabi frequencies of the red and blue lasers. We first draw, for $i\in \{795,950\}$ a realization $\phi_i(t)$ of a random process with the appropriate, experimentally measured spectral noise density $S_{\phi_i}(f)$ (see e.g.~\cite{clade2004}, page 65). We then numerically solve the appropriate dynamical equations (either the Schr\"odinger equation, either the OBEs) with Rabi frequencies $\Omega(t)=\Omega_0\exp(i\phi(t))$ for both the red and the blue lasers~\footnote{The fluctuating phase $\phi(t)$ of the blue laser is taken equal to $2\phi_{950}(t)$ as its frequency components are within the bandwidth of the cavity used for frequency doubling.}, and average over typically 100 realizations of the phase noise. In Fig.~\ref{fig:s-nu}(b), we show all individual realizations (thin red lines) and their average (solid black line) for the typical frequency noise spectra, assuming no other damping process, with a global Rabi frequency of 1~MHz and observe a slow damping of the oscillation. The influence of the phase noise is better seen when increasing it on purpose in the 950~nm laser, by applying too much gain in the feedback loop, as shown in Fig.~\ref{fig:s-nu}(a) in dashed line. In these conditions, the experimental Rabi oscillation damping is increased [disks in Fig.~\ref{fig:s-nu}(c)]. The agreement between this experiment and the parameter-free simulation validates our estimate of the phase noise and its modeling. 

\subsection{Other possible effects}
\label{other_effects}

Several other effects can in principle contribute to damping and dephasing of the Rabi oscillations. First, the Rydberg states have a finite lifetime due to spontaneous emission leading to decay to low-lying excited states and to black-body radiation transferring the atom to close-by Rydberg states~\cite{Beterov2009}. The latter has been shown to be detrimental to Rydberg-dressing experiments with many particles~\cite{Goldschmidt2016,Zeiher2016b}. We have solved the OBEs with and without including the finite lifetime of Rydberg states $n > 50$ and observed no significant differences on the timescale of a few microseconds for the single-atom Rabi oscillation. Thus, so far, this finite lifetime is not a limitation in our setup.

If the excitation beams are not perfectly centered on the trap position, the atom sits on a spatial gradient of Rabi frequency and light-shift, and thus its random thermal motion from shot to shot gives rise to dephasing. We have checked that for our experimental parameters, this effect should be negligible unless the excitation beams are strongly misaligned. Another dephasing mechanism is the shot-to-shot variation in the pulse areas of the excitation beams. We have estimated the relative fluctuations of the intensity of the pulses to be below 0.2\% rms, which does not lead to any measurable dephasing over our experimental timescales. 

Finally, stray transverse electric fields leading to mixing between different Zeeman sublevels of the targeted Rydberg state could lead to a degradation of the Rabi oscillation, as $\ket{g}$ would be coupled to several Rydberg states with different coupling strengths. However, using eight independent  electrodes under vacuum, we zero out the electric field to better than $\left \vert E \right \vert < 5~{\rm mV/cm}$ by performing Stark spectroscopy on high-$n$ Rydberg states (typically $n\sim100$). For such low values of $E$, the expected effect of stray fields is negligible. 

\begin{figure}[t!]
	\centering
	\includegraphics[width=67mm]{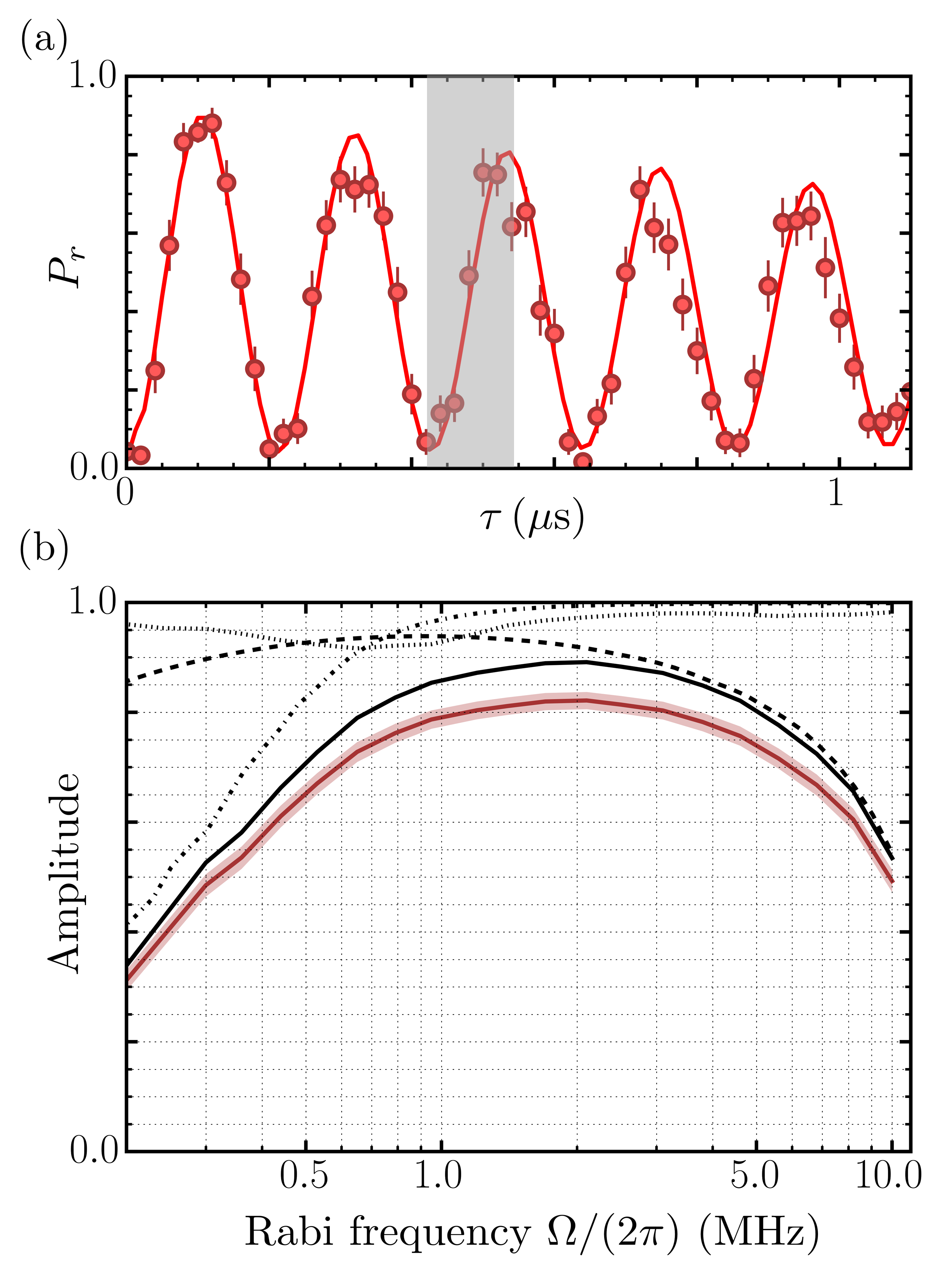}
	\caption{Including all effects in the simulation. (a) Rabi oscillation (with $n = 61$) at $\Omega/(2 \pi) = 4.8$~MHz with the red solid line showing the result of a parameter-free simulation. The gray-shaded area shows the time window used in panel (b) to quantify the damping (see text). (b) Influence of the Rabi frequency $\Omega$ (varied by changing only $\Omega_{\rm r}$, with $\Omega_b/(2 \pi) = 35$~MHz and $\Delta = 740$~MHz) on the Rabi oscillation damping. Simulations are done with typical experimental parameters (see text) for each individual source of damping: Doppler effect (dash-dotted line), spontaneous emission (dashed line), laser phase noise (dotted line) and combining all of them (solid black line) and by adding the SPAM errors (solid red line). The shading on the latter curve corresponds to s.e.m. of the Monte-Carlo simulation with 600 runs.}
	\label{fig:vary_fR}
\end{figure}

\section{Combining all effects; prospects for improvement}

Having developed a quantitative modeling of each of the experimentally relevant imperfections listed above, we can now include them all in a global simulation. All parameters (Rabi frequencies of the red and blue lasers, detuning $\Delta$, SPAM errors $(\eta, \varepsilon,\varepsilon')$, atomic temperature $T$, spectral density of the phase noise of lasers\ldots) are given their independently measured values. We draw fluctuating parameters according to their respective distributions, integrate the OBEs with these parameters, and then average over typically 600 realizations. Figure~\ref{fig:vary_fR}(a) shows a comparison between an experimental Rabi oscillation (for $\Omega/(2 \pi) = 4.8$~MHz, disks) and a parameter-free simulation (solid line) including all effects detailed in this work. The fair agreement between data and simulation allows us to explore, using our numerical simulation, how varying experimental parameters affects the coherence of the Rabi oscillation.   

As an example, we show in Fig.~\ref{fig:vary_fR}(b) how the different effects depend on the Rabi frequency $\Omega$. The simulations are performed for a fixed Rabi frequency of the blue laser $\Omega_{\rm b}/(2\pi) = 35$~MHz (typical for $n \sim 60$) and a varying red laser Rabi frequency $\Omega_{\rm r}$. To characterize the damping of the Rabi oscillation, we extract from the simulations the oscillation amplitude during the fifth half-period [gray shading in Fig.~\ref{fig:vary_fR}(a)]. When including all effects in the simulation (red solid line), we observe that the damping is minimized for $\Omega/(2\pi) \approx 2$~MHz. For $\Omega/(2 \pi) < 0.7$~MHz, the Doppler effect (dash-dotted line) is the dominant source of damping. The phase noise influence (dotted line) peaks at $\Omega/(2 \pi) = 1$~MHz as predicted from its broad maximum seen in Fig.~\ref{fig:s-nu}(a). The detrimental effect of spontaneous emission (dashed line) is minimized when $\Omega_{\rm r} = \Omega_{\rm b}$, giving $\Omega/(2 \pi) = 0.8$~MHz for our current parameters. 

One could possibly improve the coherence by increasing the power or decreasing the size of the blue beam, thus increasing $\Omega_{\rm blue}$, which would allow, for a fixed $\Omega$, to increase $\Delta$ and thus reduce spontaneous emission. However, too small a beam waist will increase sensitivity to beam pointing instability, and, more importantly, will limit the size of the tweezers arrays that can be excited homogeneously~\cite{supergaussian2016}, so ideally higher powers for the 475~nm laser would be needed. Another solution is to use the inverted excitation scheme, with intermediate state $\ket{6P_{3/2}}$, which has the combined advantage of having a longer lifetime ($\sim 120$~ns) and better coupling to Rydberg states (at 1013~nm). Using this scheme, the Harvard group obtained promising coherence times~\cite{Bernien2017, lukin_inprep}.

So far we have discussed only experimental situations in which the coherent laser drive is continuously on, which is relevant e.g. for the quantum simulation of Ising models~\cite{Labuhn2016,Bernien2017}. However, for implementing quantum gates~\cite{Saffman2016}, or for the quantum simulation of XY quantum magnets~\cite{Barredo2015}, one is interested in exciting or de-exciting selected atoms quickly and efficiently. This can be accomplished with $\pi$-pulses, in which case we are back to the problem of minimizing the decoherence of Rabi oscillations, but other schemes can be used, such as STIRAP~\cite{Raithel2005,Deiglmayr2006,Hennrich2017,stirapreview}, which has the advantage of being more forgiving in terms of fine tuning of parameters. We believe that the modeling of experimental imperfections developed in this work will be useful for finding realistic optimum parameters in those scenarios. 

\begin{acknowledgements}
We thank M. Saffman and G. Biedermann for discussions and thoughtful comments on the manuscript, M.J.~Martin for insightful suggestions during a stay in our group, and M.D. Lukin and his group for organizing a workshop at ITAMP (Cambridge, MA, USA) with a session dedicated to the issues discussed in this article. This work benefited from financial support by the EU [H2020 FET-PROACT Project RySQ], by the PALM Labex (projects QUANTICA and XYLOS) and by the R\'egion \^Ile-de-France in the framework of DIM Nano-K.
\end{acknowledgements}

\appendix

\section{A simple model for $\varepsilon'$} \label{app:epsp}

In this appendix, we derive, using a simple model, the value of the probability $\varepsilon'$ for false negatives. Let us assume that, at $t=0$, just at the end of the excitation sequence, the atom is in $\ket{r}$. We then switch on the tweezers again, and after a delay of about 10~ms (due to the time it takes to open mechanical shutters, for instance), the molasses light, in order to excite fluorescence of the atom if it is in $\ket{g}$. For the atom to emit fluorescence at the end, and thus give a false negative, it needs (i) to decay to $\ket{g}$ at some time $t$, and (ii) to be recaptured in the tweezers after having spent a time $t$ in $\ket{r}$ in the presence of the tweezers. This means that $\varepsilon'$ is given by
\begin{equation}
\varepsilon'=\int_0^\infty p_{\rm recap}(t)\,\dot{p}_g(t)\,{\rm d}t.
\label{eq:epsprime}
\end{equation}
In this expression, the quantity $p_{\rm recap}(t)$ is the probability for an atom in $\ket{g}$ to be recaptured by the (attractive) trapping potential $U({\bs r})$ of the tweezers after having experienced, for a duration $t$, the anti-trapping potential $-\alpha U({\bs r})$, $\dot{p}_g(t)$ is the time derivative of the population of $\ket{g}$ when the atom is initially in $\ket{r}$, and, finally, the upper limit of the integral can safely be replaced by $+\infty$ because the time it takes to switch on the molasses light is long compared to the timescales over which the integrand is non zero (tens of $\mu$s). The ratio $\alpha$ of the anti-trapping of $\ket{r}$ and trapping potentials of $\ket{g}$ depends on the trap laser frequency $\nu$. Assuming that the potential experienced by the Rydberg atom is essentially the ponderomotive potential exerted by the tweezers on the nearly-free valence electron, we obtain $\alpha={(\nu+\nu_0)(\nu-\nu_0)}/{\nu^2} \simeq 0.17$ (the trap laser at 852 nm corresponds to $\nu = 352$~THz, and $\nu_0 \simeq 382$~THz is the average frequency of the D lines of Rb). 

\begin{figure}[t]
	\centering
	\includegraphics[width=66mm]{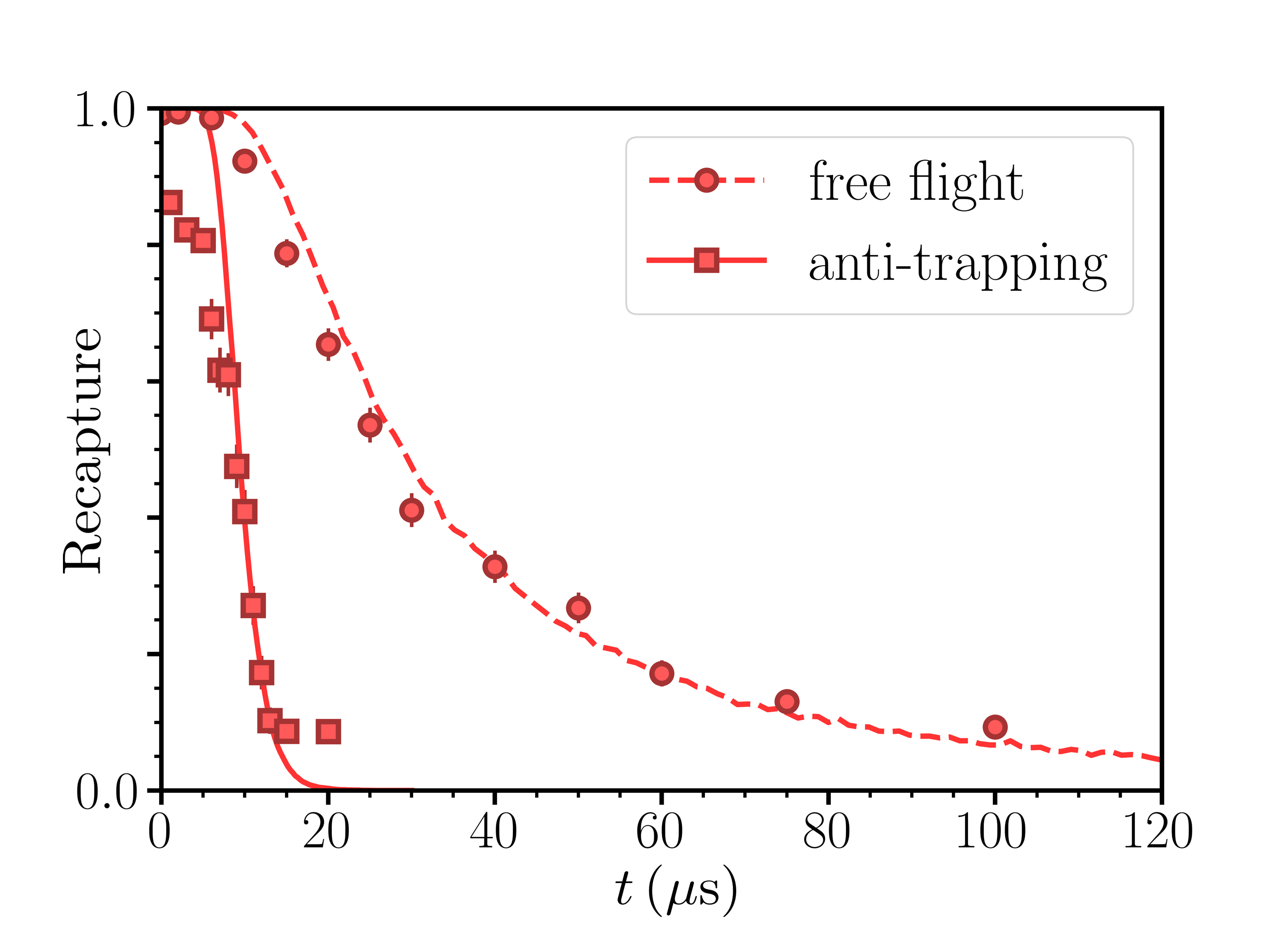}
	\caption{Recapture probability $p_{\rm recap}(t)$ including anti-trapping by the optical tweezers (solid line, red squares) and neglecting the anti-trapping (dashed line, red circles). The simulations take parameters $T=20\;\mu$K, $U_0=1$~mK, and $\alpha=0.17$ and the experiment is detailed in the text.}
	\label{fig:recapture_flight}
\end{figure}

To evaluate (\ref{eq:epsprime}), we first focus on $p_\text{recap}(t)$. We show in Fig.~\ref{fig:recapture_flight} the effect of the anti-trapping potential on a Rydberg state $\ket{r}$ compared to a free flying atom. The theoretical curves are obtained with a Monte-Carlo simulation of the classical dynamics of a particle at temperature $T = 20 \, \mu$K either in a repulsive potential $-\alpha U({\bs r})$ (solid line), either in free flight (dashed line). In a first experiment (squares), we excite an atom to $\ket{r}$ with a $\pi$-pulse, switch on the trap for a variable time $t$ and de-excite it back with a second $\pi$-pulse. The reduced contrast of the experimental data with respect to the simulation is due to the finite efficiency of Rydberg (de-)excitation. In a second experiment (circles), we simply measure the recapture probability of an atom in free flight by switching off the trap during a time $t$. Our models are in excellent agreement with the experimental data. We observe that the release and recapture measurement~\cite{Tuchendler2008} of a Rydberg atom $\ket{r}$ is strongly affected by the slight anti-trapping and drops quickly to zero after $\sim 10 \, \mu$s. This anti-trapping thus plays an important role in our detection scheme, where we need the Rydberg atom to leave the trapping region as fast as possible before it decays back to the ground state. 

Concerning $\dot{p}_g(t)$, we can write it as $\Gamma_R \exp(-t \Gamma_R)$, where $\Gamma_R$ is the rate at which a Rydberg state decays back to the ground state~\cite{Beterov2009}. At finite temperature black-body radiation does increase the depopulation rate of $\ket{r}$, but by transferring population to neighboring long-lived Rydberg states, and thus it hardly affects the rate at which $\ket{g}$ gets populated. 

For values of $n>50$, the zero-temperature lifetimes $1/\Gamma_R \propto n^3$ of $\ket{r}$ are in excess of 100~$\mu$s, while the variations of $p_{\rm recap}(t)$ occur on just a few microseconds. One can thus approximate (\ref{eq:epsprime}) by $ \varepsilon'={\Gamma_R}\,{t_{\rm recap}} $, where $t_{\rm recap}=\int_0^\infty p_{\rm recap}(t)\,{\rm d}t \simeq 10 \, \mu$s is extracted from Fig.~\ref{fig:recapture_flight} and varies only logarithmically with the atomic temperature.

\bibliographystyle{myapsrev4-1}

\begin{thebibliography}{57}%
	\makeatletter
	\providecommand \@ifxundefined [1]{%
		\@ifx{#1\undefined}
	}%
	\providecommand \@ifnum [1]{%
		\ifnum #1\expandafter \@firstoftwo
		\else \expandafter \@secondoftwo
		\fi
	}%
	\providecommand \@ifx [1]{%
		\ifx #1\expandafter \@firstoftwo
		\else \expandafter \@secondoftwo
		\fi
	}%
	\providecommand \natexlab [1]{#1}%
	\providecommand \enquote  [1]{``#1''}%
	\providecommand \bibnamefont  [1]{#1}%
	\providecommand \bibfnamefont [1]{#1}%
	\providecommand \citenamefont [1]{#1}%
	\providecommand \href@noop [0]{\@secondoftwo}%
	\providecommand \href [0]{\begingroup \@sanitize@url \@href}%
	\providecommand \@href[1]{\@@startlink{#1}\@@href}%
	\providecommand \@@href[1]{\endgroup#1\@@endlink}%
	\providecommand \@sanitize@url [0]{\catcode `\\12\catcode `\$12\catcode
		`\&12\catcode `\#12\catcode `\^12\catcode `\_12\catcode `\%12\relax}%
	\providecommand \@@startlink[1]{}%
	\providecommand \@@endlink[0]{}%
	\providecommand \url  [0]{\begingroup\@sanitize@url \@url }%
	\providecommand \@url [1]{\endgroup\@href {#1}{\urlprefix }}%
	\providecommand \urlprefix  [0]{URL }%
	\providecommand \Eprint [0]{\href }%
	\providecommand \doibase [0]{http://dx.doi.org/}%
	\providecommand \selectlanguage [0]{\@gobble}%
	\providecommand \bibinfo  [0]{\@secondoftwo}%
	\providecommand \bibfield  [0]{\@secondoftwo}%
	\providecommand \translation [1]{[#1]}%
	\providecommand \BibitemOpen [0]{}%
	\providecommand \bibitemStop [0]{}%
	\providecommand \bibitemNoStop [0]{.\EOS\space}%
	\providecommand \EOS [0]{\spacefactor3000\relax}%
	\providecommand \BibitemShut  [1]{\csname bibitem#1\endcsname}%
	\let\auto@bib@innerbib\@empty
	\bibitem [{\citenamefont {Browaeys}\ \emph {et~al.}(2016)\citenamefont
		{Browaeys}, \citenamefont {Barredo},\ and\ \citenamefont
		{Lahaye}}]{Browaeys2016}%
	\BibitemOpen
	\bibfield  {author} {\bibinfo {author} {\bibfnamefont {A.}~\bibnamefont
			{Browaeys}}, \bibinfo {author} {\bibfnamefont {D.}~\bibnamefont {Barredo}}, \
		and\ \bibinfo {author} {\bibfnamefont {T.}~\bibnamefont {Lahaye}},\ }\emph
	{Experimental investigations of dipole-dipole interactions between a few
		Rydberg atoms},\ \href {\doibase 10.1088/0953-4075/49/15/152001} {\bibfield
		{journal} {\bibinfo  {journal} {J. Phys. B}\ }\textbf {\bibinfo {volume}
			{49}},\ \bibinfo {pages} {152001} (\bibinfo {year} {2016})}\BibitemShut
	{NoStop}%
	\bibitem [{\citenamefont {Labuhn}\ \emph {et~al.}(2016)\citenamefont {Labuhn},
		\citenamefont {Barredo}, \citenamefont {Ravets}, \citenamefont
		{de~L{\'{e}}s{\'{e}}leuc}, \citenamefont {Macr{\`{i}}}, \citenamefont
		{Lahaye},\ and\ \citenamefont {Browaeys}}]{Labuhn2016}%
	\BibitemOpen
	\bibfield  {author} {\bibinfo {author} {\bibfnamefont {H.}~\bibnamefont
			{Labuhn}}, \bibinfo {author} {\bibfnamefont {D.}~\bibnamefont {Barredo}},
		\bibinfo {author} {\bibfnamefont {S.}~\bibnamefont {Ravets}}, \bibinfo
		{author} {\bibfnamefont {S.}~\bibnamefont {de~L{\'{e}}s{\'{e}}leuc}},
		\bibinfo {author} {\bibfnamefont {T.}~\bibnamefont {Macr{\`{i}}}}, \bibinfo
		{author} {\bibfnamefont {T.}~\bibnamefont {Lahaye}}, \ and\ \bibinfo {author}
		{\bibfnamefont {A.}~\bibnamefont {Browaeys}},\ }\emph {Tunable
		two-dimensional arrays of single Rydberg atoms for realizing quantum Ising
		models},\ \href {\doibase 10.1038/nature18274} {\bibfield  {journal}
		{\bibinfo  {journal} {Nature}\ }\textbf {\bibinfo {volume} {534}},\ \bibinfo
		{pages} {667} (\bibinfo {year} {2016})}\BibitemShut {NoStop}%
	\bibitem [{\citenamefont {Bernien}\ \emph {et~al.}(2017)\citenamefont
		{Bernien}, \citenamefont {Schwartz}, \citenamefont {Keesling}, \citenamefont
		{Levine}, \citenamefont {Omran}, \citenamefont {Pichler}, \citenamefont
		{Choi}, \citenamefont {Zibrov}, \citenamefont {Endres}, \citenamefont
		{Greiner}, \citenamefont {Vuleti{\'{c}}},\ and\ \citenamefont
		{Lukin}}]{Bernien2017}%
	\BibitemOpen
	\bibfield  {author} {\bibinfo {author} {\bibfnamefont {H.}~\bibnamefont
			{Bernien}}, \bibinfo {author} {\bibfnamefont {S.}~\bibnamefont {Schwartz}},
		\bibinfo {author} {\bibfnamefont {A.}~\bibnamefont {Keesling}}, \bibinfo
		{author} {\bibfnamefont {H.}~\bibnamefont {Levine}}, \bibinfo {author}
		{\bibfnamefont {A.}~\bibnamefont {Omran}}, \bibinfo {author} {\bibfnamefont
			{H.}~\bibnamefont {Pichler}}, \bibinfo {author} {\bibfnamefont
			{S.}~\bibnamefont {Choi}}, \bibinfo {author} {\bibfnamefont {A.~S.}\
			\bibnamefont {Zibrov}}, \bibinfo {author} {\bibfnamefont {M.}~\bibnamefont
			{Endres}}, \bibinfo {author} {\bibfnamefont {M.}~\bibnamefont {Greiner}},
		\bibinfo {author} {\bibfnamefont {V.}~\bibnamefont {Vuleti{\'{c}}}}, \ and\
		\bibinfo {author} {\bibfnamefont {M.~D.}\ \bibnamefont {Lukin}},\ }\emph
	{Probing many-body dynamics on a 51-atom quantum simulator},\ \href {\doibase
		10.1038/nature24622} {\bibfield  {journal} {\bibinfo  {journal} {Nature}\
		}\textbf {\bibinfo {volume} {551}},\ \bibinfo {pages} {579} (\bibinfo {year}
		{2017})}\BibitemShut {NoStop}%
	\bibitem [{\citenamefont {Lienhard}\ \emph {et~al.}(2017)\citenamefont
		{Lienhard}, \citenamefont {de~L\'es\'eleuc}, \citenamefont {Barredo},
		\citenamefont {Lahaye}, \citenamefont {Browaeys}, \citenamefont {Schuler},
		\citenamefont {Henry},\ and\ \citenamefont {Läuchli}}]{Lienhard2017}%
	\BibitemOpen
	\bibfield  {author} {\bibinfo {author} {\bibfnamefont {V.}~\bibnamefont
			{Lienhard}}, \bibinfo {author} {\bibfnamefont {S.}~\bibnamefont
			{de~L\'es\'eleuc}}, \bibinfo {author} {\bibfnamefont {D.}~\bibnamefont
			{Barredo}}, \bibinfo {author} {\bibfnamefont {T.}~\bibnamefont {Lahaye}},
		\bibinfo {author} {\bibfnamefont {A.}~\bibnamefont {Browaeys}}, \bibinfo
		{author} {\bibfnamefont {M.}~\bibnamefont {Schuler}}, \bibinfo {author}
		{\bibfnamefont {L.-P.}\ \bibnamefont {Henry}}, \ and\ \bibinfo {author}
		{\bibfnamefont {A.~M.}\ \bibnamefont {Läuchli}},\ }\emph {Observing the
		space- and time-dependent growth of correlations in dynamically tuned
		synthetic Ising antiferromagnets},\ \href {https://arxiv.org/abs/1711.01185}
	{\bibfield  {journal} {\bibinfo  {journal} {arXiv:1711.01185}\ } (\bibinfo
		{year} {2017})}\BibitemShut {NoStop}%
	\bibitem [{\citenamefont {{Kim}}\ \emph {et~al.}(2018)\citenamefont {{Kim}},
		\citenamefont {{Park}}, \citenamefont {{Kim}}, \citenamefont {{Sim}},\ and\
		\citenamefont {{Ahn}}}]{Kim2017}%
	\BibitemOpen
	\bibfield  {author} {\bibinfo {author} {\bibfnamefont {H.}~\bibnamefont
			{{Kim}}}, \bibinfo {author} {\bibfnamefont {Y.}~\bibnamefont {{Park}}},
		\bibinfo {author} {\bibfnamefont {K.}~\bibnamefont {{Kim}}}, \bibinfo
		{author} {\bibfnamefont {H.-S.}\ \bibnamefont {{Sim}}}, \ and\ \bibinfo
		{author} {\bibfnamefont {J.}~\bibnamefont {{Ahn}}},\ }\emph {{Detailed
			Balance of Thermalization dynamics in Rydberg atom quantum simulators}},\
	\href {https://arxiv.org/abs/1712.02065} {\bibfield  {journal} {\bibinfo
			{journal} {arXiv:1712.02065}\ } (\bibinfo {year} {2018})}\BibitemShut
	{NoStop}%
	\bibitem [{\citenamefont {Saffman}(2016)}]{Saffman2016}%
	\BibitemOpen
	\bibfield  {author} {\bibinfo {author} {\bibfnamefont {M.}~\bibnamefont
			{Saffman}},\ }\emph {Quantum computing with atomic qubits and Rydberg
		interactions: progress and challenges},\ \href
	{http://stacks.iop.org/0953-4075/49/i=20/a=202001} {\bibfield  {journal}
		{\bibinfo  {journal} {J. Phys. B}\ }\textbf {\bibinfo {volume} {49}},\
		\bibinfo {pages} {202001} (\bibinfo {year} {2016})}\BibitemShut {NoStop}%
	\bibitem [{\citenamefont {Wang}\ \emph {et~al.}(2015)\citenamefont {Wang},
		\citenamefont {Zhang}, \citenamefont {Corcovilos}, \citenamefont {Kumar},\
		and\ \citenamefont {Weiss}}]{Wang2015}%
	\BibitemOpen
	\bibfield  {author} {\bibinfo {author} {\bibfnamefont {Y.}~\bibnamefont
			{Wang}}, \bibinfo {author} {\bibfnamefont {X.}~\bibnamefont {Zhang}},
		\bibinfo {author} {\bibfnamefont {T.~A.}\ \bibnamefont {Corcovilos}},
		\bibinfo {author} {\bibfnamefont {A.}~\bibnamefont {Kumar}}, \ and\ \bibinfo
		{author} {\bibfnamefont {D.~S.}\ \bibnamefont {Weiss}},\ }\emph {Coherent
		Addressing of Individual Neutral Atoms in a 3D Optical Lattice},\ \href
	{\doibase 10.1103/PhysRevLett.115.043003} {\bibfield  {journal} {\bibinfo
			{journal} {Phys. Rev. Lett.}\ }\textbf {\bibinfo {volume} {115}},\ \bibinfo
		{pages} {043003} (\bibinfo {year} {2015})}\BibitemShut {NoStop}%
	\bibitem [{\citenamefont {Xia}\ \emph {et~al.}(2015)\citenamefont {Xia},
		\citenamefont {Lichtman}, \citenamefont {Maller}, \citenamefont {Carr},
		\citenamefont {Piotrowicz}, \citenamefont {Isenhower},\ and\ \citenamefont
		{Saffman}}]{Xia2015}%
	\BibitemOpen
	\bibfield  {author} {\bibinfo {author} {\bibfnamefont {T.}~\bibnamefont
			{Xia}}, \bibinfo {author} {\bibfnamefont {M.}~\bibnamefont {Lichtman}},
		\bibinfo {author} {\bibfnamefont {K.}~\bibnamefont {Maller}}, \bibinfo
		{author} {\bibfnamefont {A.~W.}\ \bibnamefont {Carr}}, \bibinfo {author}
		{\bibfnamefont {M.~J.}\ \bibnamefont {Piotrowicz}}, \bibinfo {author}
		{\bibfnamefont {L.}~\bibnamefont {Isenhower}}, \ and\ \bibinfo {author}
		{\bibfnamefont {M.}~\bibnamefont {Saffman}},\ }\emph {Randomized Benchmarking
		of Single-Qubit Gates in a 2D Array of Neutral-Atom Qubits},\ \href {\doibase
		10.1103/PhysRevLett.114.100503} {\bibfield  {journal} {\bibinfo  {journal}
			{Phys. Rev. Lett.}\ }\textbf {\bibinfo {volume} {114}},\ \bibinfo {pages}
		{100503} (\bibinfo {year} {2015})}\BibitemShut {NoStop}%
	\bibitem [{\citenamefont {Wang}\ \emph {et~al.}(2016)\citenamefont {Wang},
		\citenamefont {Kumar}, \citenamefont {Wu},\ and\ \citenamefont
		{Weiss}}]{Wang2016}%
	\BibitemOpen
	\bibfield  {author} {\bibinfo {author} {\bibfnamefont {Y.}~\bibnamefont
			{Wang}}, \bibinfo {author} {\bibfnamefont {A.}~\bibnamefont {Kumar}},
		\bibinfo {author} {\bibfnamefont {T.-Y.}\ \bibnamefont {Wu}}, \ and\ \bibinfo
		{author} {\bibfnamefont {D.~S.}\ \bibnamefont {Weiss}},\ }\emph {Single-qubit
		gates based on targeted phase shifts in a 3D neutral atom array},\ \href
	{\doibase 10.1126/science.aaf2581} {\bibfield  {journal} {\bibinfo  {journal}
			{Science}\ }\textbf {\bibinfo {volume} {352}},\ \bibinfo {pages} {1562}
		(\bibinfo {year} {2016})}\BibitemShut {NoStop}%
	\bibitem [{\citenamefont {Barredo}\ \emph {et~al.}(2016)\citenamefont
		{Barredo}, \citenamefont {de~L{\'{e}}s{\'{e}}leuc}, \citenamefont {Lienhard},
		\citenamefont {Lahaye},\ and\ \citenamefont {Browaeys}}]{Barredo2016}%
	\BibitemOpen
	\bibfield  {author} {\bibinfo {author} {\bibfnamefont {D.}~\bibnamefont
			{Barredo}}, \bibinfo {author} {\bibfnamefont {S.}~\bibnamefont
			{de~L{\'{e}}s{\'{e}}leuc}}, \bibinfo {author} {\bibfnamefont
			{V.}~\bibnamefont {Lienhard}}, \bibinfo {author} {\bibfnamefont
			{T.}~\bibnamefont {Lahaye}}, \ and\ \bibinfo {author} {\bibfnamefont
			{A.}~\bibnamefont {Browaeys}},\ }\emph {{An atom-by-atom assembler of
			defect-free arbitrary 2d atomic arrays}},\ \href {\doibase
		10.1126/science.aah3778} {\bibfield  {journal} {\bibinfo  {journal}
			{Science}\ }\textbf {\bibinfo {volume} {354}},\ \bibinfo {pages} {1021}
		(\bibinfo {year} {2016})}\BibitemShut {NoStop}%
	\bibitem [{\citenamefont {Endres}\ \emph {et~al.}(2016)\citenamefont {Endres},
		\citenamefont {Bernien}, \citenamefont {Keesling}, \citenamefont {Levine},
		\citenamefont {Anschuetz}, \citenamefont {Krajenbrink}, \citenamefont
		{Senko}, \citenamefont {Vuletic}, \citenamefont {Greiner},\ and\
		\citenamefont {Lukin}}]{Endres2016}%
	\BibitemOpen
	\bibfield  {author} {\bibinfo {author} {\bibfnamefont {M.}~\bibnamefont
			{Endres}}, \bibinfo {author} {\bibfnamefont {H.}~\bibnamefont {Bernien}},
		\bibinfo {author} {\bibfnamefont {A.}~\bibnamefont {Keesling}}, \bibinfo
		{author} {\bibfnamefont {H.}~\bibnamefont {Levine}}, \bibinfo {author}
		{\bibfnamefont {E.~R.}\ \bibnamefont {Anschuetz}}, \bibinfo {author}
		{\bibfnamefont {A.}~\bibnamefont {Krajenbrink}}, \bibinfo {author}
		{\bibfnamefont {C.}~\bibnamefont {Senko}}, \bibinfo {author} {\bibfnamefont
			{V.}~\bibnamefont {Vuletic}}, \bibinfo {author} {\bibfnamefont
			{M.}~\bibnamefont {Greiner}}, \ and\ \bibinfo {author} {\bibfnamefont
			{M.~D.}\ \bibnamefont {Lukin}},\ }\emph {Atom-by-atom assembly of defect-free
		one-dimensional cold atom arrays},\ \href {\doibase 10.1126/science.aah3752}
	{\bibfield  {journal} {\bibinfo  {journal} {Science}\ }\textbf {\bibinfo
			{volume} {354}},\ \bibinfo {pages} {1024} (\bibinfo {year}
		{2016})}\BibitemShut {NoStop}%
	\bibitem [{\citenamefont {Barredo}\ \emph {et~al.}(2017)\citenamefont
		{Barredo}, \citenamefont {Lienhard}, \citenamefont {de~L\'es\'eleuc},
		\citenamefont {Lahaye},\ and\ \citenamefont {Browaeys}}]{Barredo2017}%
	\BibitemOpen
	\bibfield  {author} {\bibinfo {author} {\bibfnamefont {D.}~\bibnamefont
			{Barredo}}, \bibinfo {author} {\bibfnamefont {V.}~\bibnamefont {Lienhard}},
		\bibinfo {author} {\bibfnamefont {S.}~\bibnamefont {de~L\'es\'eleuc}},
		\bibinfo {author} {\bibfnamefont {T.}~\bibnamefont {Lahaye}}, \ and\ \bibinfo
		{author} {\bibfnamefont {A.}~\bibnamefont {Browaeys}},\ }\emph {Synthetic
		three-dimensional atomic structures assembled atom by atom},\ \href
	{https://arxiv.org/abs/1712.02727} {\bibfield  {journal} {\bibinfo  {journal}
			{arXiv:1712.02727}\ } (\bibinfo {year} {2017})}\BibitemShut {NoStop}%
	\bibitem [{\citenamefont {Deiglmayr}\ \emph {et~al.}(2006)\citenamefont
		{Deiglmayr}, \citenamefont {Reetz-Lamour}, \citenamefont {Amthor},
		\citenamefont {Westermann}, \citenamefont {de~Oliveira},\ and\ \citenamefont
		{Weidem\"uller}}]{Deiglmayr2006}%
	\BibitemOpen
	\bibfield  {author} {\bibinfo {author} {\bibfnamefont {J.}~\bibnamefont
			{Deiglmayr}}, \bibinfo {author} {\bibfnamefont {M.}~\bibnamefont
			{Reetz-Lamour}}, \bibinfo {author} {\bibfnamefont {T.}~\bibnamefont
			{Amthor}}, \bibinfo {author} {\bibfnamefont {S.}~\bibnamefont {Westermann}},
		\bibinfo {author} {\bibfnamefont {A.}~\bibnamefont {de~Oliveira}}, \ and\
		\bibinfo {author} {\bibfnamefont {M.}~\bibnamefont {Weidem\"uller}},\ }\emph
	{Coherent excitation of Rydberg atoms in an ultracold gas},\ \href {\doibase
		10.1016/j.optcom.2006.02.058} {\bibfield  {journal} {\bibinfo  {journal}
			{Optics Comm.}\ }\textbf {\bibinfo {volume} {264}},\ \bibinfo {pages} {293}
		(\bibinfo {year} {2006})}\BibitemShut {NoStop}%
	\bibitem [{\citenamefont {Reetz-Lamour}\ \emph {et~al.}(2008)\citenamefont
		{Reetz-Lamour}, \citenamefont {Amthor}, \citenamefont {Deiglmayr},\ and\
		\citenamefont {Weidem\"uller}}]{Reetz-Lamour2008}%
	\BibitemOpen
	\bibfield  {author} {\bibinfo {author} {\bibfnamefont {M.}~\bibnamefont
			{Reetz-Lamour}}, \bibinfo {author} {\bibfnamefont {T.}~\bibnamefont
			{Amthor}}, \bibinfo {author} {\bibfnamefont {J.}~\bibnamefont {Deiglmayr}}, \
		and\ \bibinfo {author} {\bibfnamefont {M.}~\bibnamefont {Weidem\"uller}},\
	}\emph {Rabi Oscillations and Excitation Trapping in the Coherent Excitation
	of a Mesoscopic Frozen Rydberg Gas},\ \href {\doibase
	10.1103/PhysRevLett.100.253001} {\bibfield  {journal} {\bibinfo  {journal}
		{Phys. Rev. Lett.}\ }\textbf {\bibinfo {volume} {100}},\ \bibinfo {pages}
	{253001} (\bibinfo {year} {2008})}\BibitemShut {NoStop}%
\bibitem [{\citenamefont {Johnson}\ \emph {et~al.}(2008)\citenamefont
	{Johnson}, \citenamefont {Urban}, \citenamefont {Henage}, \citenamefont
	{Isenhower}, \citenamefont {Yavuz}, \citenamefont {Walker},\ and\
	\citenamefont {Saffman}}]{Johnson2008}%
\BibitemOpen
\bibfield  {author} {\bibinfo {author} {\bibfnamefont {T.~A.}\ \bibnamefont
		{Johnson}}, \bibinfo {author} {\bibfnamefont {E.}~\bibnamefont {Urban}},
	\bibinfo {author} {\bibfnamefont {T.}~\bibnamefont {Henage}}, \bibinfo
	{author} {\bibfnamefont {L.}~\bibnamefont {Isenhower}}, \bibinfo {author}
	{\bibfnamefont {D.~D.}\ \bibnamefont {Yavuz}}, \bibinfo {author}
	{\bibfnamefont {T.~G.}\ \bibnamefont {Walker}}, \ and\ \bibinfo {author}
	{\bibfnamefont {M.}~\bibnamefont {Saffman}},\ }\emph {Rabi Oscillations
	between Ground and Rydberg States with Dipole-Dipole Atomic Interactions},\
\href {\doibase 10.1103/PhysRevLett.100.113003} {\bibfield  {journal}
	{\bibinfo  {journal} {Phys. Rev. Lett.}\ }\textbf {\bibinfo {volume} {100}},\
	\bibinfo {pages} {113003} (\bibinfo {year} {2008})}\BibitemShut {NoStop}%
\bibitem [{\citenamefont {Zuo}\ \emph {et~al.}(2009)\citenamefont {Zuo},
	\citenamefont {Fukusen}, \citenamefont {Tamaki}, \citenamefont {Watanabe},
	\citenamefont {Nakagawa},\ and\ \citenamefont {Nakagawa}}]{Zuo2009}%
\BibitemOpen
\bibfield  {author} {\bibinfo {author} {\bibfnamefont {Z.}~\bibnamefont
		{Zuo}}, \bibinfo {author} {\bibfnamefont {M.}~\bibnamefont {Fukusen}},
	\bibinfo {author} {\bibfnamefont {Y.}~\bibnamefont {Tamaki}}, \bibinfo
	{author} {\bibfnamefont {T.}~\bibnamefont {Watanabe}}, \bibinfo {author}
	{\bibfnamefont {Y.}~\bibnamefont {Nakagawa}}, \ and\ \bibinfo {author}
	{\bibfnamefont {K.}~\bibnamefont {Nakagawa}},\ }\emph {Single atom Rydberg
	excitation in a small dipole trap},\ \href {\doibase 10.1364/OE.17.022898}
{\bibfield  {journal} {\bibinfo  {journal} {Opt. Express}\ }\textbf {\bibinfo
		{volume} {17}},\ \bibinfo {pages} {22898} (\bibinfo {year}
	{2009})}\BibitemShut {NoStop}%
\bibitem [{\citenamefont {Miroshnychenko}\ \emph {et~al.}(2010)\citenamefont
	{Miroshnychenko}, \citenamefont {Ga\"etan}, \citenamefont {Evellin},
	\citenamefont {Grangier}, \citenamefont {Comparat}, \citenamefont {Pillet},
	\citenamefont {Wilk},\ and\ \citenamefont {Browaeys}}]{Miroshnychenko2010}%
\BibitemOpen
\bibfield  {author} {\bibinfo {author} {\bibfnamefont {Y.}~\bibnamefont
		{Miroshnychenko}}, \bibinfo {author} {\bibfnamefont {A.}~\bibnamefont
		{Ga\"etan}}, \bibinfo {author} {\bibfnamefont {C.}~\bibnamefont {Evellin}},
	\bibinfo {author} {\bibfnamefont {P.}~\bibnamefont {Grangier}}, \bibinfo
	{author} {\bibfnamefont {D.}~\bibnamefont {Comparat}}, \bibinfo {author}
	{\bibfnamefont {P.}~\bibnamefont {Pillet}}, \bibinfo {author} {\bibfnamefont
		{T.}~\bibnamefont {Wilk}}, \ and\ \bibinfo {author} {\bibfnamefont
		{A.}~\bibnamefont {Browaeys}},\ }\emph {Coherent excitation of a single atom
	to a Rydberg state},\ \href {\doibase 10.1103/PhysRevA.82.013405} {\bibfield
	{journal} {\bibinfo  {journal} {Phys. Rev. A}\ }\textbf {\bibinfo {volume}
		{82}},\ \bibinfo {pages} {013405} (\bibinfo {year} {2010})}\BibitemShut
{NoStop}%
\bibitem [{\citenamefont {Hankin}\ \emph {et~al.}(2014)\citenamefont {Hankin},
	\citenamefont {Jau}, \citenamefont {Parazzoli}, \citenamefont {Chou},
	\citenamefont {Armstrong}, \citenamefont {Landahl},\ and\ \citenamefont
	{Biedermann}}]{Biedermann2014}%
\BibitemOpen
\bibfield  {author} {\bibinfo {author} {\bibfnamefont {A.~M.}\ \bibnamefont
		{Hankin}}, \bibinfo {author} {\bibfnamefont {Y.-Y.}\ \bibnamefont {Jau}},
	\bibinfo {author} {\bibfnamefont {L.~P.}\ \bibnamefont {Parazzoli}}, \bibinfo
	{author} {\bibfnamefont {C.~W.}\ \bibnamefont {Chou}}, \bibinfo {author}
	{\bibfnamefont {D.~J.}\ \bibnamefont {Armstrong}}, \bibinfo {author}
	{\bibfnamefont {A.~J.}\ \bibnamefont {Landahl}}, \ and\ \bibinfo {author}
	{\bibfnamefont {G.~W.}\ \bibnamefont {Biedermann}},\ }\emph {Two-atom Rydberg
	blockade using direct 6$S$ to $nP$ excitation},\ \href {\doibase
	10.1103/PhysRevA.89.033416} {\bibfield  {journal} {\bibinfo  {journal} {Phys.
			Rev. A}\ }\textbf {\bibinfo {volume} {89}},\ \bibinfo {pages} {033416}
	(\bibinfo {year} {2014})}\BibitemShut {NoStop}%
\bibitem [{\citenamefont {Dudin}\ \emph {et~al.}(2012)\citenamefont {Dudin},
	\citenamefont {Li}, \citenamefont {Bariani},\ and\ \citenamefont
	{Kuzmich}}]{Dudin2012}%
\BibitemOpen
\bibfield  {author} {\bibinfo {author} {\bibfnamefont {Y.~O.}\ \bibnamefont
		{Dudin}}, \bibinfo {author} {\bibfnamefont {L.}~\bibnamefont {Li}}, \bibinfo
	{author} {\bibfnamefont {F.}~\bibnamefont {Bariani}}, \ and\ \bibinfo
	{author} {\bibfnamefont {A.}~\bibnamefont {Kuzmich}},\ }\emph {Observation of
	coherent many-body Rabi oscillations},\ \href {\doibase 10.1038/nphys2413}
{\bibfield  {journal} {\bibinfo  {journal} {Nat. Phys.}\ }\textbf {\bibinfo
		{volume} {8}},\ \bibinfo {pages} {790} (\bibinfo {year} {2012})}\BibitemShut
{NoStop}%
\bibitem [{\citenamefont {Ebert}\ \emph {et~al.}(2015)\citenamefont {Ebert},
	\citenamefont {Kwon}, \citenamefont {Walker},\ and\ \citenamefont
	{Saffman}}]{Ebert2015}%
\BibitemOpen
\bibfield  {author} {\bibinfo {author} {\bibfnamefont {M.}~\bibnamefont
		{Ebert}}, \bibinfo {author} {\bibfnamefont {M.}~\bibnamefont {Kwon}},
	\bibinfo {author} {\bibfnamefont {T.~G.}\ \bibnamefont {Walker}}, \ and\
	\bibinfo {author} {\bibfnamefont {M.}~\bibnamefont {Saffman}},\ }\emph
{Coherence and Rydberg Blockade of Atomic Ensemble Qubits},\ \href {\doibase
	10.1103/PhysRevLett.115.093601} {\bibfield  {journal} {\bibinfo  {journal}
		{Phys. Rev. Lett.}\ }\textbf {\bibinfo {volume} {115}},\ \bibinfo {pages}
	{093601} (\bibinfo {year} {2015})}\BibitemShut {NoStop}%
\bibitem [{\citenamefont {Zeiher}\ \emph {et~al.}(2015)\citenamefont {Zeiher},
	\citenamefont {Schau\ss{}}, \citenamefont {Hild}, \citenamefont {Macr\`{\i}},
	\citenamefont {Bloch},\ and\ \citenamefont {Gross}}]{Zeiher2016}%
\BibitemOpen
\bibfield  {author} {\bibinfo {author} {\bibfnamefont {J.}~\bibnamefont
		{Zeiher}}, \bibinfo {author} {\bibfnamefont {P.}~\bibnamefont {Schau\ss{}}},
	\bibinfo {author} {\bibfnamefont {S.}~\bibnamefont {Hild}}, \bibinfo {author}
	{\bibfnamefont {T.}~\bibnamefont {Macr\`{\i}}}, \bibinfo {author}
	{\bibfnamefont {I.}~\bibnamefont {Bloch}}, \ and\ \bibinfo {author}
	{\bibfnamefont {C.}~\bibnamefont {Gross}},\ }\emph {Microscopic
	Characterization of Scalable Coherent Rydberg Superatoms},\ \href {\doibase
	10.1103/PhysRevX.5.031015} {\bibfield  {journal} {\bibinfo  {journal} {Phys.
			Rev. X}\ }\textbf {\bibinfo {volume} {5}},\ \bibinfo {pages} {031015}
	(\bibinfo {year} {2015})}\BibitemShut {NoStop}%
\bibitem [{\citenamefont {Schau\ss}\ \emph {et~al.}(2012)\citenamefont
	{Schau\ss}, \citenamefont {Cheneau}, \citenamefont {Endres}, \citenamefont
	{Fukuhara}, \citenamefont {Hild}, \citenamefont {Omran}, \citenamefont
	{Pohl}, \citenamefont {Gross}, \citenamefont {Kuhr},\ and\ \citenamefont
	{Bloch}}]{Schauss2012}%
\BibitemOpen
\bibfield  {author} {\bibinfo {author} {\bibfnamefont {P.}~\bibnamefont
		{Schau\ss}}, \bibinfo {author} {\bibfnamefont {M.}~\bibnamefont {Cheneau}},
	\bibinfo {author} {\bibfnamefont {M.}~\bibnamefont {Endres}}, \bibinfo
	{author} {\bibfnamefont {T.}~\bibnamefont {Fukuhara}}, \bibinfo {author}
	{\bibfnamefont {S.}~\bibnamefont {Hild}}, \bibinfo {author} {\bibfnamefont
		{A.}~\bibnamefont {Omran}}, \bibinfo {author} {\bibfnamefont
		{T.}~\bibnamefont {Pohl}}, \bibinfo {author} {\bibfnamefont {C.}~\bibnamefont
		{Gross}}, \bibinfo {author} {\bibfnamefont {S.}~\bibnamefont {Kuhr}}, \ and\
	\bibinfo {author} {\bibfnamefont {I.}~\bibnamefont {Bloch}},\ }\emph
{Observation of mesoscopic crystalline structures in a two-dimensional
	{R}ydberg gas},\ \href {\doibase 10.1038/nature11596} {\bibfield  {journal}
	{\bibinfo  {journal} {Nature}\ }\textbf {\bibinfo {volume} {491}},\ \bibinfo
	{pages} {87} (\bibinfo {year} {2012})}\BibitemShut {NoStop}%
\bibitem [{\citenamefont {Schau\ss{}}\ \emph {et~al.}(2015)\citenamefont
	{Schau\ss{}}, \citenamefont {Zeiher}, \citenamefont {Fukuhara}, \citenamefont
	{Hild}, \citenamefont {Cheneau}, \citenamefont {Macri}, \citenamefont {Pohl},
	\citenamefont {Bloch},\ and\ \citenamefont {Gross}}]{Schauss2015}%
\BibitemOpen
\bibfield  {author} {\bibinfo {author} {\bibfnamefont {P.}~\bibnamefont
		{Schau\ss{}}}, \bibinfo {author} {\bibfnamefont {J.}~\bibnamefont {Zeiher}},
	\bibinfo {author} {\bibfnamefont {T.}~\bibnamefont {Fukuhara}}, \bibinfo
	{author} {\bibfnamefont {S.}~\bibnamefont {Hild}}, \bibinfo {author}
	{\bibfnamefont {M.}~\bibnamefont {Cheneau}}, \bibinfo {author} {\bibfnamefont
		{T.}~\bibnamefont {Macri}}, \bibinfo {author} {\bibfnamefont
		{T.}~\bibnamefont {Pohl}}, \bibinfo {author} {\bibfnamefont {I.}~\bibnamefont
		{Bloch}}, \ and\ \bibinfo {author} {\bibfnamefont {C.}~\bibnamefont
		{Gross}},\ }\emph {{Crystallization in Ising quantum magnets}},\ \href
{\doibase 10.1126/science.1258351} {\bibfield  {journal} {\bibinfo  {journal}
		{Science}\ }\textbf {\bibinfo {volume} {347}},\ \bibinfo {pages} {1455}
	(\bibinfo {year} {2015})}\BibitemShut {NoStop}%
\bibitem [{\citenamefont {Guardado-Sanchez}\ \emph {et~al.}(2017)\citenamefont
	{Guardado-Sanchez}, \citenamefont {Brown}, \citenamefont {Mitra},
	\citenamefont {Devakul}, \citenamefont {Huse}, \citenamefont {Schauss},\ and\
	\citenamefont {Bakr}}]{Guardado2017}%
\BibitemOpen
\bibfield  {author} {\bibinfo {author} {\bibfnamefont {E.}~\bibnamefont
		{Guardado-Sanchez}}, \bibinfo {author} {\bibfnamefont {P.~T.}\ \bibnamefont
		{Brown}}, \bibinfo {author} {\bibfnamefont {D.}~\bibnamefont {Mitra}},
	\bibinfo {author} {\bibfnamefont {T.}~\bibnamefont {Devakul}}, \bibinfo
	{author} {\bibfnamefont {D.~A.}\ \bibnamefont {Huse}}, \bibinfo {author}
	{\bibfnamefont {P.}~\bibnamefont {Schauss}}, \ and\ \bibinfo {author}
	{\bibfnamefont {W.~S.}\ \bibnamefont {Bakr}},\ }\emph {Probing quench
	dynamics across a quantum phase transition into a 2D Ising antiferromagnet},\
\href {https://arxiv.org/abs/1711.00887} {\bibfield  {journal} {\bibinfo
		{journal} {arXiv:1711.00887}\ } (\bibinfo {year} {2017})}\BibitemShut
{NoStop}%
\bibitem [{\citenamefont {Maller}\ \emph {et~al.}(2015)\citenamefont {Maller},
	\citenamefont {Lichtman}, \citenamefont {Xia}, \citenamefont {Sun},
	\citenamefont {Piotrowicz}, \citenamefont {Carr}, \citenamefont {Isenhower},\
	and\ \citenamefont {Saffman}}]{Maller2015}%
\BibitemOpen
\bibfield  {author} {\bibinfo {author} {\bibfnamefont {K.~M.}\ \bibnamefont
		{Maller}}, \bibinfo {author} {\bibfnamefont {M.~T.}\ \bibnamefont
		{Lichtman}}, \bibinfo {author} {\bibfnamefont {T.}~\bibnamefont {Xia}},
	\bibinfo {author} {\bibfnamefont {Y.}~\bibnamefont {Sun}}, \bibinfo {author}
	{\bibfnamefont {M.~J.}\ \bibnamefont {Piotrowicz}}, \bibinfo {author}
	{\bibfnamefont {A.~W.}\ \bibnamefont {Carr}}, \bibinfo {author}
	{\bibfnamefont {L.}~\bibnamefont {Isenhower}}, \ and\ \bibinfo {author}
	{\bibfnamefont {M.}~\bibnamefont {Saffman}},\ }\emph {Rydberg-blockade
	controlled-not gate and entanglement in a two-dimensional array of
	neutral-atom qubits},\ \href {\doibase 10.1103/PhysRevA.92.022336} {\bibfield
	{journal} {\bibinfo  {journal} {Phys. Rev. A}\ }\textbf {\bibinfo {volume}
		{92}},\ \bibinfo {pages} {022336} (\bibinfo {year} {2015})}\BibitemShut
{NoStop}%
\bibitem [{\citenamefont {Jau}\ \emph {et~al.}(2015)\citenamefont {Jau},
	\citenamefont {Hankin}, \citenamefont {Keating}, \citenamefont {Deutsch},\
	and\ \citenamefont {Biedermann}}]{Jau2015}%
\BibitemOpen
\bibfield  {author} {\bibinfo {author} {\bibfnamefont {Y.-Y.}\ \bibnamefont
		{Jau}}, \bibinfo {author} {\bibfnamefont {A.~M.}\ \bibnamefont {Hankin}},
	\bibinfo {author} {\bibfnamefont {T.}~\bibnamefont {Keating}}, \bibinfo
	{author} {\bibfnamefont {I.~H.}\ \bibnamefont {Deutsch}}, \ and\ \bibinfo
	{author} {\bibfnamefont {G.~W.}\ \bibnamefont {Biedermann}},\ }\emph
{Entangling atomic spins with a Rydberg-dressed spin-flip blockade},\ \href
{\doibase 10.1038/nphys3487} {\bibfield  {journal} {\bibinfo  {journal} {Nat.
			Phys.}\ }\textbf {\bibinfo {volume} {12}},\ \bibinfo {pages} {71} (\bibinfo
	{year} {2015})}\BibitemShut {NoStop}%
\bibitem [{\citenamefont {Xia}\ \emph {et~al.}(2013)\citenamefont {Xia},
	\citenamefont {Zhang},\ and\ \citenamefont {Saffman}}]{Xia2013}%
\BibitemOpen
\bibfield  {author} {\bibinfo {author} {\bibfnamefont {T.}~\bibnamefont
		{Xia}}, \bibinfo {author} {\bibfnamefont {X.~L.}\ \bibnamefont {Zhang}}, \
	and\ \bibinfo {author} {\bibfnamefont {M.}~\bibnamefont {Saffman}},\ }\emph
{Analysis of a controlled phase gate using circular Rydberg states},\ \href
{\doibase 10.1103/PhysRevA.88.062337} {\bibfield  {journal} {\bibinfo
		{journal} {Phys. Rev. A}\ }\textbf {\bibinfo {volume} {88}},\ \bibinfo
	{pages} {062337} (\bibinfo {year} {2013})}\BibitemShut {NoStop}%
\bibitem [{\citenamefont {Petrosyan}\ \emph {et~al.}(2017)\citenamefont
	{Petrosyan}, \citenamefont {Motzoi}, \citenamefont {Saffman},\ and\
	\citenamefont {M\o{}lmer}}]{Petrosyan2017}%
\BibitemOpen
\bibfield  {author} {\bibinfo {author} {\bibfnamefont {D.}~\bibnamefont
		{Petrosyan}}, \bibinfo {author} {\bibfnamefont {F.}~\bibnamefont {Motzoi}},
	\bibinfo {author} {\bibfnamefont {M.}~\bibnamefont {Saffman}}, \ and\
	\bibinfo {author} {\bibfnamefont {K.}~\bibnamefont {M\o{}lmer}},\ }\emph
{High-fidelity Rydberg quantum gate via a two-atom dark state},\ \href
{\doibase 10.1103/PhysRevA.96.042306} {\bibfield  {journal} {\bibinfo
		{journal} {Phys. Rev. A}\ }\textbf {\bibinfo {volume} {96}},\ \bibinfo
	{pages} {042306} (\bibinfo {year} {2017})}\BibitemShut {NoStop}%
\bibitem [{\citenamefont {Monz}\ \emph {et~al.}(2011)\citenamefont {Monz},
	\citenamefont {Schindler}, \citenamefont {Barreiro}, \citenamefont {Chwalla},
	\citenamefont {Nigg}, \citenamefont {Coish}, \citenamefont {Harlander},
	\citenamefont {H\"ansel}, \citenamefont {Hennrich},\ and\ \citenamefont
	{Blatt}}]{Monz2011}%
\BibitemOpen
\bibfield  {author} {\bibinfo {author} {\bibfnamefont {T.}~\bibnamefont
		{Monz}}, \bibinfo {author} {\bibfnamefont {P.}~\bibnamefont {Schindler}},
	\bibinfo {author} {\bibfnamefont {J.~T.}\ \bibnamefont {Barreiro}}, \bibinfo
	{author} {\bibfnamefont {M.}~\bibnamefont {Chwalla}}, \bibinfo {author}
	{\bibfnamefont {D.}~\bibnamefont {Nigg}}, \bibinfo {author} {\bibfnamefont
		{W.~A.}\ \bibnamefont {Coish}}, \bibinfo {author} {\bibfnamefont
		{M.}~\bibnamefont {Harlander}}, \bibinfo {author} {\bibfnamefont
		{W.}~\bibnamefont {H\"ansel}}, \bibinfo {author} {\bibfnamefont
		{M.}~\bibnamefont {Hennrich}}, \ and\ \bibinfo {author} {\bibfnamefont
		{R.}~\bibnamefont {Blatt}},\ }\emph {14-Qubit Entanglement: Creation and
	Coherence},\ \href {\doibase 10.1103/PhysRevLett.106.130506} {\bibfield
	{journal} {\bibinfo  {journal} {Phys. Rev. Lett.}\ }\textbf {\bibinfo
		{volume} {106}},\ \bibinfo {pages} {130506} (\bibinfo {year}
	{2011})}\BibitemShut {NoStop}%
\bibitem [{\citenamefont {Ballance}\ \emph {et~al.}(2016)\citenamefont
	{Ballance}, \citenamefont {Harty}, \citenamefont {Linke}, \citenamefont
	{Sepiol},\ and\ \citenamefont {Lucas}}]{Ballance2016}%
\BibitemOpen
\bibfield  {author} {\bibinfo {author} {\bibfnamefont {C.~J.}\ \bibnamefont
		{Ballance}}, \bibinfo {author} {\bibfnamefont {T.~P.}\ \bibnamefont {Harty}},
	\bibinfo {author} {\bibfnamefont {N.~M.}\ \bibnamefont {Linke}}, \bibinfo
	{author} {\bibfnamefont {M.~A.}\ \bibnamefont {Sepiol}}, \ and\ \bibinfo
	{author} {\bibfnamefont {D.~M.}\ \bibnamefont {Lucas}},\ }\emph
{High-Fidelity Quantum Logic Gates Using Trapped-Ion Hyperfine Qubits},\
\href {\doibase 10.1103/PhysRevLett.117.060504} {\bibfield  {journal}
	{\bibinfo  {journal} {Phys. Rev. Lett.}\ }\textbf {\bibinfo {volume} {117}},\
	\bibinfo {pages} {060504} (\bibinfo {year} {2016})}\BibitemShut {NoStop}%
\bibitem [{\citenamefont {Chow}\ \emph {et~al.}(2012)\citenamefont {Chow},
	\citenamefont {Gambetta}, \citenamefont {C\'orcoles}, \citenamefont {Merkel},
	\citenamefont {Smolin}, \citenamefont {Rigetti}, \citenamefont {Poletto},
	\citenamefont {Keefe}, \citenamefont {Rothwell}, \citenamefont {Rozen},
	\citenamefont {Ketchen},\ and\ \citenamefont {Steffen}}]{Chow2012}%
\BibitemOpen
\bibfield  {author} {\bibinfo {author} {\bibfnamefont {J.~M.}\ \bibnamefont
		{Chow}}, \bibinfo {author} {\bibfnamefont {J.~M.}\ \bibnamefont {Gambetta}},
	\bibinfo {author} {\bibfnamefont {A.~D.}\ \bibnamefont {C\'orcoles}},
	\bibinfo {author} {\bibfnamefont {S.~T.}\ \bibnamefont {Merkel}}, \bibinfo
	{author} {\bibfnamefont {J.~A.}\ \bibnamefont {Smolin}}, \bibinfo {author}
	{\bibfnamefont {C.}~\bibnamefont {Rigetti}}, \bibinfo {author} {\bibfnamefont
		{S.}~\bibnamefont {Poletto}}, \bibinfo {author} {\bibfnamefont {G.~A.}\
		\bibnamefont {Keefe}}, \bibinfo {author} {\bibfnamefont {M.~B.}\ \bibnamefont
		{Rothwell}}, \bibinfo {author} {\bibfnamefont {J.~R.}\ \bibnamefont {Rozen}},
	\bibinfo {author} {\bibfnamefont {M.~B.}\ \bibnamefont {Ketchen}}, \ and\
	\bibinfo {author} {\bibfnamefont {M.}~\bibnamefont {Steffen}},\ }\emph
{Universal Quantum Gate Set Approaching Fault-Tolerant Thresholds with
	Superconducting Qubits},\ \href {\doibase 10.1103/PhysRevLett.109.060501}
{\bibfield  {journal} {\bibinfo  {journal} {Phys. Rev. Lett.}\ }\textbf
	{\bibinfo {volume} {109}},\ \bibinfo {pages} {060501} (\bibinfo {year}
	{2012})}\BibitemShut {NoStop}%
\bibitem [{\citenamefont {Barends~{\emph{et al.}}}(2014)}]{Martinis2014_short}%
\BibitemOpen
\bibfield  {author} {\bibinfo {author} {\bibfnamefont {R.}~\bibnamefont
		{Barends~{\emph{et al.}}}},\ }\emph {Superconducting quantum circuits at the
	surface code threshold for fault tolerance},\ \href {\doibase
	10.1038/nature13171} {\bibfield  {journal} {\bibinfo  {journal} {Nature}\
	}\textbf {\bibinfo {volume} {508}},\ \bibinfo {pages} {500} (\bibinfo {year}
	{2014})}\BibitemShut {NoStop}%
\bibitem [{\citenamefont {Song~{\emph{et al.}}}(2017)}]{Pan2017_short}%
\BibitemOpen
\bibfield  {author} {\bibinfo {author} {\bibfnamefont {C.}~\bibnamefont
		{Song~{\emph{et al.}}}},\ }\emph {10-Qubit Entanglement and Parallel Logic
	Operations with a Superconducting Circuit},\ \href {\doibase
	10.1103/PhysRevLett.119.180511} {\bibfield  {journal} {\bibinfo  {journal}
		{Phys. Rev. Lett.}\ }\textbf {\bibinfo {volume} {119}},\ \bibinfo {pages}
	{180511} (\bibinfo {year} {2017})}\BibitemShut {NoStop}%
\bibitem [{\citenamefont {Zhang}\ \emph {et~al.}(2010)\citenamefont {Zhang},
	\citenamefont {Isenhower}, \citenamefont {Gill}, \citenamefont {Walker},\
	and\ \citenamefont {Saffman}}]{Zhang2010}%
\BibitemOpen
\bibfield  {author} {\bibinfo {author} {\bibfnamefont {X.~L.}\ \bibnamefont
		{Zhang}}, \bibinfo {author} {\bibfnamefont {L.}~\bibnamefont {Isenhower}},
	\bibinfo {author} {\bibfnamefont {A.~T.}\ \bibnamefont {Gill}}, \bibinfo
	{author} {\bibfnamefont {T.~G.}\ \bibnamefont {Walker}}, \ and\ \bibinfo
	{author} {\bibfnamefont {M.}~\bibnamefont {Saffman}},\ }\emph {Deterministic
	entanglement of two neutral atoms via Rydberg blockade},\ \href {\doibase
	10.1103/PhysRevA.82.030306} {\bibfield  {journal} {\bibinfo  {journal} {Phys.
			Rev. A}\ }\textbf {\bibinfo {volume} {82}},\ \bibinfo {pages} {030306}
	(\bibinfo {year} {2010})}\BibitemShut {NoStop}%
\bibitem [{pri()}]{private}%
\BibitemOpen
\href@noop {} {}\bibinfo {note} {Private communications by S. Schwartz, M.J.
	Martin, and M. Saffman}\BibitemShut {NoStop}%
\bibitem [{\citenamefont {Tuchendler}\ \emph {et~al.}(2008)\citenamefont
	{Tuchendler}, \citenamefont {Lance}, \citenamefont {Browaeys}, \citenamefont
	{Sortais},\ and\ \citenamefont {Grangier}}]{Tuchendler2008}%
\BibitemOpen
\bibfield  {author} {\bibinfo {author} {\bibfnamefont {C.}~\bibnamefont
		{Tuchendler}}, \bibinfo {author} {\bibfnamefont {A.~M.}\ \bibnamefont
		{Lance}}, \bibinfo {author} {\bibfnamefont {A.}~\bibnamefont {Browaeys}},
	\bibinfo {author} {\bibfnamefont {Y.~R.~P.}\ \bibnamefont {Sortais}}, \ and\
	\bibinfo {author} {\bibfnamefont {P.}~\bibnamefont {Grangier}},\ }\emph
{Energy distribution and cooling of a single atom in an optical tweezer},\
\href {\doibase 10.1103/PhysRevA.78.033425} {\bibfield  {journal} {\bibinfo
		{journal} {Phys. Rev. A}\ }\textbf {\bibinfo {volume} {78}},\ \bibinfo
	{pages} {033425} (\bibinfo {year} {2008})}\BibitemShut {NoStop}%
\bibitem [{\citenamefont {Higgins}\ \emph {et~al.}(2017)\citenamefont
	{Higgins}, \citenamefont {Pokorny}, \citenamefont {Zhang}, \citenamefont
	{Bodart},\ and\ \citenamefont {Hennrich}}]{Hennrich2017}%
\BibitemOpen
\bibfield  {author} {\bibinfo {author} {\bibfnamefont {G.}~\bibnamefont
		{Higgins}}, \bibinfo {author} {\bibfnamefont {F.}~\bibnamefont {Pokorny}},
	\bibinfo {author} {\bibfnamefont {C.}~\bibnamefont {Zhang}}, \bibinfo
	{author} {\bibfnamefont {Q.}~\bibnamefont {Bodart}}, \ and\ \bibinfo {author}
	{\bibfnamefont {M.}~\bibnamefont {Hennrich}},\ }\emph {{Coherent Control of a
		Single Trapped Rydberg Ion}},\ \href {\doibase
	10.1103/PhysRevLett.119.220501} {\bibfield  {journal} {\bibinfo  {journal}
		{Phys. Rev. Lett.}\ }\textbf {\bibinfo {volume} {119}},\ \bibinfo {pages}
	{220501} (\bibinfo {year} {2017})}\BibitemShut {NoStop}%
\bibitem [{\citenamefont {Shen}\ and\ \citenamefont {Duan}(2012)}]{Duan2012}%
\BibitemOpen
\bibfield  {author} {\bibinfo {author} {\bibfnamefont {C.}~\bibnamefont
		{Shen}}\ and\ \bibinfo {author} {\bibfnamefont {L.-M.}\ \bibnamefont
		{Duan}},\ }\emph {Correcting detection errors in quantum state engineering
	through data processing},\ \href
{http://stacks.iop.org/1367-2630/14/i=5/a=053053} {\bibfield  {journal}
	{\bibinfo  {journal} {New J. Phys.}\ }\textbf {\bibinfo {volume} {14}},\
	\bibinfo {pages} {053053} (\bibinfo {year} {2012})}\BibitemShut {NoStop}%
\bibitem [{\citenamefont {Barredo}\ \emph {et~al.}(2015)\citenamefont
	{Barredo}, \citenamefont {Labuhn}, \citenamefont {Ravets}, \citenamefont
	{Lahaye}, \citenamefont {Browaeys},\ and\ \citenamefont
	{Adams}}]{Barredo2015}%
\BibitemOpen
\bibfield  {author} {\bibinfo {author} {\bibfnamefont {D.}~\bibnamefont
		{Barredo}}, \bibinfo {author} {\bibfnamefont {H.}~\bibnamefont {Labuhn}},
	\bibinfo {author} {\bibfnamefont {S.}~\bibnamefont {Ravets}}, \bibinfo
	{author} {\bibfnamefont {T.}~\bibnamefont {Lahaye}}, \bibinfo {author}
	{\bibfnamefont {A.}~\bibnamefont {Browaeys}}, \ and\ \bibinfo {author}
	{\bibfnamefont {C.~S.}\ \bibnamefont {Adams}},\ }\emph {Coherent Excitation
	Transfer in a Spin Chain of Three {R}ydberg Atoms},\ \href {\doibase
	10.1103/PhysRevLett.114.113002} {\bibfield  {journal} {\bibinfo  {journal}
		{Phys. Rev. Lett.}\ }\textbf {\bibinfo {volume} {114}},\ \bibinfo {pages}
	{113002} (\bibinfo {year} {2015})}\BibitemShut {NoStop}%
\bibitem [{\citenamefont {Martinez-Dorantes}\ \emph {et~al.}(2017)\citenamefont
	{Martinez-Dorantes}, \citenamefont {Alt}, \citenamefont {Gallego},
	\citenamefont {Ghosh}, \citenamefont {Ratschbacher},\ and\ \citenamefont
	{Meschede}}]{Martinez-Dorantes2017b}%
\BibitemOpen
\bibfield  {author} {\bibinfo {author} {\bibfnamefont {M.}~\bibnamefont
		{Martinez-Dorantes}}, \bibinfo {author} {\bibfnamefont {W.}~\bibnamefont
		{Alt}}, \bibinfo {author} {\bibfnamefont {J.}~\bibnamefont {Gallego}},
	\bibinfo {author} {\bibfnamefont {S.}~\bibnamefont {Ghosh}}, \bibinfo
	{author} {\bibfnamefont {L.}~\bibnamefont {Ratschbacher}}, \ and\ \bibinfo
	{author} {\bibfnamefont {D.}~\bibnamefont {Meschede}},\ }\emph
{State-dependent fluorescence of neutral atoms in optical potentials},\ \href
{https://arxiv.org/abs/1710.07964} {\bibfield  {journal} {\bibinfo  {journal}
		{arXiv:1710.07964}\ } (\bibinfo {year} {2017})}\BibitemShut {NoStop}%
\bibitem [{Note1()}]{Note1}%
\BibitemOpen
\bibinfo {note} {A laser beam resonant on the ${\delimiter "026A30C } F=2
	{\delimiter "526930B } {\rightarrow } {\delimiter "026A30C } F' = 3
	{\delimiter "526930B }$ cycling transition at 780~nm is shone during $4
	\protect \tmspace +\thinmuskip {.1667em} \mu $s (in the presence of a
	repumper beam). Atoms in any hyperfine and Zeeman levels of $5S_{1/2}$ are
	ejected with a probability better than $99.5\%$, while atoms in ${\delimiter
		"026A30C } r {\delimiter "526930B }$ are not affected. A similar technique
	was used for the detection of Rydberg states on a dark background in~\cite
	{Schauss2012}.}\BibitemShut {Stop}%
\bibitem [{\citenamefont {Zhang}\ \emph {et~al.}(2011)\citenamefont {Zhang},
	\citenamefont {Robicheaux},\ and\ \citenamefont {Saffman}}]{Zhang2011}%
\BibitemOpen
\bibfield  {author} {\bibinfo {author} {\bibfnamefont {S.}~\bibnamefont
		{Zhang}}, \bibinfo {author} {\bibfnamefont {F.}~\bibnamefont {Robicheaux}}, \
	and\ \bibinfo {author} {\bibfnamefont {M.}~\bibnamefont {Saffman}},\ }\emph
{Magic-wavelength optical traps for Rydberg atoms},\ \href {\doibase
	10.1103/PhysRevA.84.043408} {\bibfield  {journal} {\bibinfo  {journal} {Phys.
			Rev. A}\ }\textbf {\bibinfo {volume} {84}},\ \bibinfo {pages} {043408}
	(\bibinfo {year} {2011})}\BibitemShut {NoStop}%
\bibitem [{\citenamefont {Ryabtsev}\ \emph {et~al.}(2011)\citenamefont
	{Ryabtsev}, \citenamefont {Beterov}, \citenamefont {Tretyakov}, \citenamefont
	{Entin},\ and\ \citenamefont {Yakshina}}]{Ryabtsev2011}%
\BibitemOpen
\bibfield  {author} {\bibinfo {author} {\bibfnamefont {I.~I.}\ \bibnamefont
		{Ryabtsev}}, \bibinfo {author} {\bibfnamefont {I.~I.}\ \bibnamefont
		{Beterov}}, \bibinfo {author} {\bibfnamefont {D.~B.}\ \bibnamefont
		{Tretyakov}}, \bibinfo {author} {\bibfnamefont {V.~M.}\ \bibnamefont
		{Entin}}, \ and\ \bibinfo {author} {\bibfnamefont {E.~A.}\ \bibnamefont
		{Yakshina}},\ }\emph {Doppler- and recoil-free laser excitation of Rydberg
	states via three-photon transitions},\ \href {\doibase
	10.1103/PhysRevA.84.053409} {\bibfield  {journal} {\bibinfo  {journal} {Phys.
			Rev. A}\ }\textbf {\bibinfo {volume} {84}},\ \bibinfo {pages} {053409}
	(\bibinfo {year} {2011})}\BibitemShut {NoStop}%
\bibitem [{\citenamefont {Kaufman}\ \emph {et~al.}(2012)\citenamefont
	{Kaufman}, \citenamefont {Lester},\ and\ \citenamefont
	{Regal}}]{Kaufman2012}%
\BibitemOpen
\bibfield  {author} {\bibinfo {author} {\bibfnamefont {A.~M.}\ \bibnamefont
		{Kaufman}}, \bibinfo {author} {\bibfnamefont {B.~J.}\ \bibnamefont {Lester}},
	\ and\ \bibinfo {author} {\bibfnamefont {C.~A.}\ \bibnamefont {Regal}},\
}\emph {Cooling a Single Atom in an Optical Tweezer to Its Quantum Ground
State},\ \href {\doibase 10.1103/PhysRevX.2.041014} {\bibfield  {journal}
{\bibinfo  {journal} {Phys. Rev. X}\ }\textbf {\bibinfo {volume} {2}},\
\bibinfo {pages} {041014} (\bibinfo {year} {2012})}\BibitemShut {NoStop}%
\bibitem [{\citenamefont {Thompson}\ \emph {et~al.}(2013)\citenamefont
	{Thompson}, \citenamefont {Tiecke}, \citenamefont {Zibrov}, \citenamefont
	{Vuleti\ifmmode~\acute{c}\else \'{c}\fi{}},\ and\ \citenamefont
	{Lukin}}]{Thompson2013}%
\BibitemOpen
\bibfield  {author} {\bibinfo {author} {\bibfnamefont {J.~D.}\ \bibnamefont
		{Thompson}}, \bibinfo {author} {\bibfnamefont {T.~G.}\ \bibnamefont
		{Tiecke}}, \bibinfo {author} {\bibfnamefont {A.~S.}\ \bibnamefont {Zibrov}},
	\bibinfo {author} {\bibfnamefont {V.}~\bibnamefont
		{Vuleti\ifmmode~\acute{c}\else \'{c}\fi{}}}, \ and\ \bibinfo {author}
	{\bibfnamefont {M.~D.}\ \bibnamefont {Lukin}},\ }\emph {Coherence and Raman
	Sideband Cooling of a Single Atom in an Optical Tweezer},\ \href {\doibase
	10.1103/PhysRevLett.110.133001} {\bibfield  {journal} {\bibinfo  {journal}
		{Phys. Rev. Lett.}\ }\textbf {\bibinfo {volume} {110}},\ \bibinfo {pages}
	{133001} (\bibinfo {year} {2013})}\BibitemShut {NoStop}%
\bibitem [{phd()}]{phdtarallo}%
\BibitemOpen
\href@noop {} {}\bibinfo {note} {M.G. Tarallo, PhD thesis, Universit\`a degli
	studi di Pisa, (2009).}\BibitemShut {Stop}%
\bibitem [{rie()}]{riehle2004}%
\BibitemOpen
\href@noop {} {}\bibinfo {note} {F. Riehle, \emph{Frequency standards, basics
		and applications}, Wiley-VCH, Weinheim (2004).}\BibitemShut {Stop}%
\bibitem [{mar()}]{martin}%
\BibitemOpen
\href@noop {} {}\bibinfo {note} {M.J. Martin, PhD thesis, University of
	Colorado (2013).}\BibitemShut {Stop}%
\bibitem [{cla()}]{clade2004}%
\BibitemOpen
\href@noop {} {}\bibinfo {note} {P. Clad\'e, PhD thesis, Universit\'e Paris 6
	(2004).}\BibitemShut {Stop}%
\bibitem [{Note2()}]{Note2}%
\BibitemOpen
\bibinfo {note} {The fluctuating phase $\phi (t)$ of the blue laser is taken
	equal to $2\phi _{950}(t)$ as its frequency components are within the
	bandwidth of the cavity used for frequency doubling.}\BibitemShut {Stop}%
\bibitem [{\citenamefont {Beterov}\ \emph {et~al.}(2009)\citenamefont
	{Beterov}, \citenamefont {Ryabtsev}, \citenamefont {Tretyakov},\ and\
	\citenamefont {Entin}}]{Beterov2009}%
\BibitemOpen
\bibfield  {author} {\bibinfo {author} {\bibfnamefont {I.~I.}\ \bibnamefont
		{Beterov}}, \bibinfo {author} {\bibfnamefont {I.~I.}\ \bibnamefont
		{Ryabtsev}}, \bibinfo {author} {\bibfnamefont {D.~B.}\ \bibnamefont
		{Tretyakov}}, \ and\ \bibinfo {author} {\bibfnamefont {V.~M.}\ \bibnamefont
		{Entin}},\ }\emph {Quasiclassical calculations of blackbody-radiation-induced
	depopulation rates and effective lifetimes of Rydberg $nS$, $nP$, and $nD$
	alkali-metal atoms with $n\ensuremath{\le}80$},\ \href {\doibase
	10.1103/PhysRevA.79.052504} {\bibfield  {journal} {\bibinfo  {journal} {Phys.
			Rev. A}\ }\textbf {\bibinfo {volume} {79}},\ \bibinfo {pages} {052504}
	(\bibinfo {year} {2009})}\BibitemShut {NoStop}%
\bibitem [{\citenamefont {Goldschmidt}\ \emph {et~al.}(2016)\citenamefont
	{Goldschmidt}, \citenamefont {Boulier}, \citenamefont {Brown}, \citenamefont
	{Koller}, \citenamefont {Young}, \citenamefont {Gorshkov}, \citenamefont
	{Rolston},\ and\ \citenamefont {Porto}}]{Goldschmidt2016}%
\BibitemOpen
\bibfield  {author} {\bibinfo {author} {\bibfnamefont {E.~A.}\ \bibnamefont
		{Goldschmidt}}, \bibinfo {author} {\bibfnamefont {T.}~\bibnamefont
		{Boulier}}, \bibinfo {author} {\bibfnamefont {R.~C.}\ \bibnamefont {Brown}},
	\bibinfo {author} {\bibfnamefont {S.~B.}\ \bibnamefont {Koller}}, \bibinfo
	{author} {\bibfnamefont {J.~T.}\ \bibnamefont {Young}}, \bibinfo {author}
	{\bibfnamefont {A.~V.}\ \bibnamefont {Gorshkov}}, \bibinfo {author}
	{\bibfnamefont {S.~L.}\ \bibnamefont {Rolston}}, \ and\ \bibinfo {author}
	{\bibfnamefont {J.~V.}\ \bibnamefont {Porto}},\ }\emph {Anomalous Broadening
	in Driven Dissipative Rydberg Systems},\ \href {\doibase
	10.1103/PhysRevLett.116.113001} {\bibfield  {journal} {\bibinfo  {journal}
		{Phys. Rev. Lett.}\ }\textbf {\bibinfo {volume} {116}},\ \bibinfo {pages}
	{113001} (\bibinfo {year} {2016})}\BibitemShut {NoStop}%
\bibitem [{\citenamefont {Zeiher}\ \emph {et~al.}(2016)\citenamefont {Zeiher},
	\citenamefont {van Bijnen}, \citenamefont {Schau\ss{}}, \citenamefont {Hild},
	\citenamefont {Choi}, \citenamefont {Pohl}, \citenamefont {Bloch},\ and\
	\citenamefont {Gross}}]{Zeiher2016b}%
\BibitemOpen
\bibfield  {author} {\bibinfo {author} {\bibfnamefont {J.}~\bibnamefont
		{Zeiher}}, \bibinfo {author} {\bibfnamefont {R.}~\bibnamefont {van Bijnen}},
	\bibinfo {author} {\bibfnamefont {P.}~\bibnamefont {Schau\ss{}}}, \bibinfo
	{author} {\bibfnamefont {S.}~\bibnamefont {Hild}}, \bibinfo {author}
	{\bibfnamefont {J.-y.}\ \bibnamefont {Choi}}, \bibinfo {author}
	{\bibfnamefont {T.}~\bibnamefont {Pohl}}, \bibinfo {author} {\bibfnamefont
		{I.}~\bibnamefont {Bloch}}, \ and\ \bibinfo {author} {\bibfnamefont
		{C.}~\bibnamefont {Gross}},\ }\emph {Many-body interferometry of a
	Rydberg-dressed spin lattice},\ \href {\doibase 10.1038/nphys3835} {\bibfield
	{journal} {\bibinfo  {journal} {Nat. Phys.}\ }\textbf {\bibinfo {volume}
		{12}},\ \bibinfo {pages} {1095} (\bibinfo {year} {2016})}\BibitemShut
{NoStop}%
\bibitem [{\citenamefont {Gillen-Christandl}\ \emph {et~al.}(2016)\citenamefont
	{Gillen-Christandl}, \citenamefont {Gillen}, \citenamefont {Piotrowicz},\
	and\ \citenamefont {Saffman}}]{supergaussian2016}%
\BibitemOpen
\bibfield  {author} {\bibinfo {author} {\bibfnamefont {K.}~\bibnamefont
		{Gillen-Christandl}}, \bibinfo {author} {\bibfnamefont {G.~D.}\ \bibnamefont
		{Gillen}}, \bibinfo {author} {\bibfnamefont {M.~J.}\ \bibnamefont
		{Piotrowicz}}, \ and\ \bibinfo {author} {\bibfnamefont {M.}~\bibnamefont
		{Saffman}},\ }\emph {Comparison of Gaussian and super Gaussian laser beams
	for addressing atomic qubits},\ \href {\doibase 10.1007/s00340-016-6407-y}
{\bibfield  {journal} {\bibinfo  {journal} {Applied Physics B}\ }\textbf
	{\bibinfo {volume} {122}},\ \bibinfo {pages} {131} (\bibinfo {year}
	{2016})}\BibitemShut {NoStop}%
\bibitem [{\citenamefont {{H. Levine, A. Keesling, A. Omran, H. Bernien, S.
			Schwartz, A. S. Zibrov, M. Endres, M. Greiner, V. Vuleti\`c, M. D.
			Lukin}}(2018)}]{lukin_inprep}%
\BibitemOpen
\bibfield  {author} {\bibinfo {author} {\bibnamefont {{H. Levine, A.
				Keesling, A. Omran, H. Bernien, S. Schwartz, A. S. Zibrov, M. Endres, M.
				Greiner, V. Vuleti\`c, M. D. Lukin}}},\ }\emph {Improved coherent control of
	Rydberg atom qubits and entangled pairs},\ \href@noop {} {\bibfield
	{journal} {\bibinfo  {journal} {in preparation}\ } (\bibinfo {year}
	{2018})}\BibitemShut {NoStop}%
\bibitem [{\citenamefont {Cubel}\ \emph {et~al.}(2005)\citenamefont {Cubel},
	\citenamefont {Teo}, \citenamefont {Malinovsky}, \citenamefont {Guest},
	\citenamefont {Reinhard}, \citenamefont {Knuffman}, \citenamefont {Berman},\
	and\ \citenamefont {Raithel}}]{Raithel2005}%
\BibitemOpen
\bibfield  {author} {\bibinfo {author} {\bibfnamefont {T.}~\bibnamefont
		{Cubel}}, \bibinfo {author} {\bibfnamefont {B.~K.}\ \bibnamefont {Teo}},
	\bibinfo {author} {\bibfnamefont {V.~S.}\ \bibnamefont {Malinovsky}},
	\bibinfo {author} {\bibfnamefont {J.~R.}\ \bibnamefont {Guest}}, \bibinfo
	{author} {\bibfnamefont {A.}~\bibnamefont {Reinhard}}, \bibinfo {author}
	{\bibfnamefont {B.}~\bibnamefont {Knuffman}}, \bibinfo {author}
	{\bibfnamefont {P.~R.}\ \bibnamefont {Berman}}, \ and\ \bibinfo {author}
	{\bibfnamefont {G.}~\bibnamefont {Raithel}},\ }\emph {Coherent population
	transfer of ground-state atoms into Rydberg states},\ \href {\doibase
	10.1103/PhysRevA.72.023405} {\bibfield  {journal} {\bibinfo  {journal} {Phys.
			Rev. A}\ }\textbf {\bibinfo {volume} {72}},\ \bibinfo {pages} {023405}
	(\bibinfo {year} {2005})}\BibitemShut {NoStop}%
\bibitem [{\citenamefont {Vitanov}\ \emph {et~al.}(2017)\citenamefont
	{Vitanov}, \citenamefont {Rangelov}, \citenamefont {Shore},\ and\
	\citenamefont {Bergmann}}]{stirapreview}%
\BibitemOpen
\bibfield  {author} {\bibinfo {author} {\bibfnamefont {N.~V.}\ \bibnamefont
		{Vitanov}}, \bibinfo {author} {\bibfnamefont {A.~A.}\ \bibnamefont
		{Rangelov}}, \bibinfo {author} {\bibfnamefont {B.~W.}\ \bibnamefont {Shore}},
	\ and\ \bibinfo {author} {\bibfnamefont {K.}~\bibnamefont {Bergmann}},\
}\emph {Stimulated Raman adiabatic passage in physics, chemistry, and
beyond},\ \href {\doibase 10.1103/RevModPhys.89.015006} {\bibfield  {journal}
{\bibinfo  {journal} {Rev. Mod. Phys.}\ }\textbf {\bibinfo {volume} {89}},\
\bibinfo {pages} {015006} (\bibinfo {year} {2017})}\BibitemShut {NoStop}%
\end{thebibliography}
%

\end{document}